\documentclass[
	reprint,
showkeys,
	aps,
	prd,
	longbibliography,
nofootinbib, 
onecolumn
]{revtex4-2}

\usepackage[utf8]{inputenc}
\usepackage{tabularx,ragged2e}
\newcolumntype{Y}{>{\centering\arraybackslash}X} 
\usepackage{amsmath}
\usepackage{amssymb}
\usepackage{upgreek}
\usepackage{physics}
\usepackage{slashed}

\usepackage{hhline}
\usepackage{dsfont}

\usepackage[hyperref]{xcolor}
	\definecolor{goethe-blau}{cmyk}{1.0,0.2,0.0,0.4}
	\definecolor{hellgrau}{cmyk}{0.04,0.04,0.05,0.02}
	\definecolor{sandgrau}{cmyk}{0.12,0.09,0.13,0.0}
	\definecolor{dunkelgrau}{cmyk}{0.25,0.25,0.30,0.75}
	\definecolor{emo-rot}{cmyk}{0.04,1.0,0.8,0.07}
	\definecolor{purple}{cmyk}{0.08,1.0,0.3,0.36}
	\definecolor{senfgelb}{cmyk}{0.01,0.25,1.0,0.05}
	\definecolor{gruen}{cmyk}{0.62,0.4,0.87,0.09}
	\definecolor{magenta}{cmyk}{0.08,0.86,0.12,0.12}
	\definecolor{orange}{cmyk}{0.0,0.7,1.0,0.04}
	\definecolor{sonnengelb}{cmyk}{0.0,0.12,0.95,0.0}
	\definecolor{helles-gruen}{cmyk}{0.4,0.17,0.81,0.07}
	\definecolor{lichtblau}{cmyk}{0.8,0.0,0.06,0.04}

\usepackage[
	colorlinks,
	pdfpagelabels,
	breaklinks,
	pdfstartview=FitH,
	bookmarksopen=true,
	bookmarksnumbered=true,
	bookmarksopenlevel=2,
	plainpages=false,
	hypertexnames=false,
	citecolor=emo-rot,
	linkcolor=goethe-blau,
	urlcolor=purple,
	pdftitle={Spatially oscillating correlation functions in $\left(2+1\right)$-dimensional four-fermion models: The mixing of scalar and vector modes at finite density}, 
	pdfauthor={Marc Winstel},
	pdfkeywords={}
]{hyperref}

\usepackage[capitalise]{cleveref}
\usepackage{multirow}
\usepackage{xstring}
\usepackage{graphicx}	
\usepackage[acronym]{glossaries-extra}
\setabbreviationstyle[acronym]{long-short}

\newacronym{ff}{FF}{four-fermion}
\newacronym[plural=QFTs, longplural=quantum field theories]{qft}{QFT}{quantum field theory}
\newacronym{qcd}{QCD}{Quantum Chromodynamics}
\newacronym{cep}{CEP}{critical end point}
\newacronym{lp}{LP}{Lifshitz point}
\newacronym{njl}{NJL}{Nambu-Jona-Lasinio}
\newacronym{cnjl}{PSFF}{complete Lorentz-(pseudo)scalar four-fermion}
\newacronym{qm}{QM}{quark-meson}
\newacronym{gn}{GN}{Gross-Neveu}
\newacronym{chign}{$\chi$HGN}{chiral Heisenberg-Gross-Neveu}
\newacronym[longplural=inhomogeneous phases]{ip}{IP}{inhomogeneous phase}
\newacronym{hbp}{HBP}{homogeneous broken phase}
\newacronym{sp}{SP}{symmetry-restored phase}
\newacronym{uv}{UV}{ultra-violet}
\glssetcategoryattribute{acronym}{nohyperfirst}{true}

\providecommand{\Rcite}[1]{\begingroup
	\def\tempx{0}\StrCount{#1}{,}[\tempx]\ifnum\tempx > 0 
	Refs.~\else
	Ref.~\fi
	\endgroup
	\cite{#1}}

\newcommand{\Q}{\mathrm{Q}}

\newcommand{\ii}{\ensuremath{\mathrm{i}}}
\newcommand{\e}{\ensuremath{\mathrm{e}}}

\newcommand{\U}{\ensuremath{\mathrm{U}}}

\DeclareMathOperator*{\argmax}{arg\,max}
\DeclareMathOperator*{\argmin}{arg\,min}

\newcommand{\pot}{\ensuremath{U}}
\newcommand{\HomPot}{\ensuremath{\Hom{\pot}}}
\newcommand{\FD}{\ensuremath{n_F}}
\newcommand{\AFD}{\ensuremath{n_{\bar{F}}}}

\renewcommand{\Re}{\operatorname{Re}}
\renewcommand{\Im}{\operatorname{Im}}

\newcommand{\stV}[1]{\ensuremath{#1}}
\newcommand{\xstV}{\ensuremath{\stV{x}}}

\newcommand{\kmax}{\ensuremath{k_{\mathrm{max}}}}

\newcommand{\Hom}[1]{\ensuremath{\bar{#1}}}
\newcommand{\Hs}{\ensuremath{\Hom{\sigma}}}
\newcommand{\Hw}{\ensuremath{\Hom{\omega}_3}}

\newcommand{\Prop}{\mathrm{S}}
\newcommand{\minHs}{\ensuremath{\Hom{\Sigma}}}
\newcommand{\minHw}{\ensuremath{\Hom{\Omega}_3}}
\newcommand{\minHn}{\ensuremath{\Hom{N}}}
\newcommand{\Hn}{\ensuremath{\Hom{n}}}
\newcommand{\muSh}{\ensuremath{\Hom{\mu}}}
\newcommand{\so}{\ensuremath{\minHs_0}}

\newcommand{\auxf}{\ensuremath{\phi}}
\newcommand{\dbf}{\ensuremath{\varphi}}

\newcommand{\minauxf}{\ensuremath{\Phi}}

\newcommand{\ve}[1]{\ensuremath{\vec{#1}}}
\newcommand{\vauxf}{\ensuremath{\ve{\auxf}}}
\newcommand{\vdbf}{\ensuremath{\ve{\dbf}}}

\newcommand{\vminauxf}{\ensuremath{\ve{\minauxf}}}

\newcommand{\Hminauxf}{\ensuremath{\Hom{\minauxf}}}

\newcommand{\pt}[1]{\ensuremath{\delta #1}}
\newcommand{\ptauxf}{\ensuremath{\pt{\auxf}}}

\newcommand{\ptdbf}{\ensuremath{\pt{\dbf}}}

\newcommand{\ft}[1]{\ensuremath{\tilde{#1}}}
\newcommand{\ptftauxf}{\ensuremath{\pt{\ft{\auxf}}}}

\newcommand{\pauli}{\ensuremath{\tau}}
\newcommand{\vpauli}{\ensuremath{\vec{\pauli}}}

\newcommand{\cm}{\ensuremath{c}}
\newcommand{\intchannelSet}{\ensuremath{C}}

\newcommand{\dr}{\ensuremath{\mathrm{d}}}
\newcommand{\coupling}{\ensuremath{\lambda}}

\renewcommand{\S}{\mathcal{S}}
\newcommand{\seff}{\mathcal{S}_{\text{eff}}}

\newcommand{\N}{\ensuremath{N}}

\begin{document}
	
	\preprint{APS/123-QED}
	
	\title{
		Spatially oscillating correlation functions in $\left(2+1\right)$-dimensional four-fermion models: The mixing of scalar and vector modes at finite density
	}

	\author{Marc Winstel}
	\email{winstel@itp.uni-frankfurt.de}
	
	\affiliation{
		Institut für Theoretische Physik, Goethe-Universität,
		\\
		Max-von-Laue-Straße 1, D-60438 Frankfurt am Main, Germany.
	}
	
	\date{\today}
	
	\begin{abstract}
		In this work, we demonstrate that the mixing of scalar and vector condensates produces spatially oscillating, but exponentially damped correlation functions in fermionic theories at finite density and temperature.
		We find a regime exhibiting this oscillatory behavior in a Gross-Neveu-type model that also features vector interactions within the mean-field approximation.
		The existence of this regime aligns with expectations based on symmetry arguments, that are also applicable to QCD at finite baryon density.
		We compute the phase diagram including both homogeneous phases and regions with spatially oscillating, exponentially damped correlation functions at finite temperature and chemical potential for different strengths of the vector coupling.
		Furthermore, we find that inhomogeneous condensates are disfavored compared to homogeneous ones akin to previous findings without vector interactions \cite{Pannullo:2023one}.
		We show that our results are valid for a broad class of $\left(2+1\right)$-dimensional models with local four-fermion interactions. 
	\end{abstract} 

	\maketitle
	
	\tableofcontents
	
	\section{Introduction}
	\Glspl{qft} with fermionic four-point interactions -- so-called \gls{ff} models -- and Yukawa models are often used as qualitative descriptions of fermionic matter in various branches of physics, mostly high-energy and condensed matter physics \cite{Asakawa:1989bq, Klevansky:1992qe,Scavenius:2000qd, Hands:2002mr, Sadzikowski:2002iy, Hands:2004uv, Schaefer:2004en, Drut:2008rg, Hiller:2008eh, Gatto:2012sp, Roscher:2013cma, Klimenko:2012qi, Braguta:2013klm, Andersen:2014xxa, Hands:2015lza, Gracey:2018cff, Gracey:2018qba, Lang:2018csk, Farias:2016let, Khunjua:2017khh, Knorr:2017yze, Tripolt:2017zgc, Braguta:2019pxt, Ferrer:2019zfp, Lakaschus:2020caq, Lopes:2021tro, Attanasio:2021jkk, Gyory:2022hnv,Attanasio:2022fsr,Attanasio:2023fte, Ferrer:2023xvl,Ayala:2023cnt, Podo:2023ute, Pitsinigkos:2023xee, TabatabaeeMehr:2023tpt}.
	An important application is the phenomenological low-energy description of chiral symmetry breaking in \gls{qcd} at finite temperature $T$ and chemical potential $\mu$.
	Although originally designed as models for the strong interaction between nucleons, a common example for \gls{qcd}-inspired models are the \gls{njl} and the \gls{qm} model, which describe the interaction of fermions through the exchange of light mesons, such as, e.g., pions \cite{PhysRev.122.345, PhysRev.124.246, Gell-Mann:1960mvl}.
	When formulated in the chiral limit, i.e., without a bare mass term for the fermions, these models typically feature an \gls{hbp} with a chiral condensate $\langle \bar \psi \psi \rangle \neq 0 $ acting as a dynamically generated mass term in the vacuum. 
	Above a critical temperature, a second-order phase transition from the \gls{hbp} to the \gls{sp} occurs where chiral symmetry is restored -- in consistency with the chiral cross-over of \gls{qcd} with physical quark masses\footnote{In the case of formulating these models with a non-vanishing bare quark mass, the chiral symmetry is also only approximate and a crossover is observed instead of a second-order phase transition. The explicit symmetry breaking can then be compensated at high temperature and chemical potential in a super restoration phase boundary, as found in \Rcite{Inagaki:2023ktz}.} \cite{Borsanyi:2010bp}.
	The effective model description is of particular relevance for phenomenological predictions at non-vanishing density, since \gls{qcd} suffers from complex Boltzmann weights appearing in the partition function at $\mu \neq 0$ \cite{deForcrand:2009zkb}.
	The complex weights in the partition function significantly complicate first-principle calculations, e.g, using lattice field theory.
	Therefore, the development of techniques with access to the \gls{qcd} phase diagram in the $(\mu,T)$ plane is an active field of research \cite{Fischer:2018sdj, Fu:2019hdw, Bernhardt:2022mnx,Motta:2023xwm, Bernhardt:2023hpr, Ihssen:2023xlp, Motta:2023pks, Zorbach:2024zjx}. 
	
	Thus, except for regions of the \gls{qcd} phase diagram at $\mu = 0$ and small $\mu /T$, theorists have relied on effective models for the study of the phase diagram of strongly-interacting matter at intermediate densities and temperatures, see, e.g., \Rcite{Asakawa:1989bq, Scavenius:2000qd}.
	Model studies are often carried out in the mean-field approximation where bosonic quantum fluctuations are suppressed -- in order to reduce the complexity of the calculations.
	With respect to the chiral phase transition, computations assuming homogeneous ground states typically discuss the scenario of a first-order phase transition including a \gls{cep} at non-vanishing $\mu$ -- in consistency with investigations of \gls{qcd} using functional methods \cite{Fischer:2018sdj, Fu:2019hdw}.
	When allowing for inhomogeneous condensates in these model calculations the first order phase transition region is replaced by the alternative scenario of a so-called \gls{ip} -- a phase where the chiral condensate is a function of the spatial coordinates, i.e, $\langle \bar \psi \psi \rangle = f(\mathbf{x})$, and translational symmetry is spontaneously broken (see \Rcite{Buballa:2014tba} for a review).
	A phenomenon expected to be closely related to the \gls{ip} is the so-called moat regime -- where a negative bosonic wave-function renormalization is obtained.
	Accordingly, a modified dispersion relation with a minimal energy at a non-vanishing momentum is obtained \cite{Pisarski:2021qof}, which is the reason that the moat regime is also called a precursor phenomenon of an \gls{ip}, see also \Rcite{Koenigstein:2021llr} for an example of a $\left(1+1\right)$-dimensional model featuring a moat regime which is present in large parts of the \gls{ip} and the \gls{sp}.
	In general, spatially oscillatory behavior of quantities that are related to particles with the moat dispersion relation should be favored within a moat regime.
	Scenarios, which are alternatives to translational symmetry breaking in an \gls{ip}, include a liquid crystal-like behavior \cite{Kolehmainen:1982jn, Fradkin:2013sab, Hidaka:2015xza, Lee:2015bva, Akerlund:2016myr, Pisarski:2018bct}, where correlations are oscillatory and of quasi-long range order with polynomial suppression, and a so-called `quantum pion liquid' with oscillatory, but exponentially suppressed two-point correlation functions \cite{Schindler:2019ugo, Pisarski:2020dnx, Schindler:2021otf, Schindler:2021cke}.
	Both of these scenarios are related to some sort of disordering of the \gls{ip} through bosonic quantum fluctuations.
	The observation of the moat regime in a recent Functional Renormalization Group study \cite{Fu:2019hdw} as well as a few other, albeit limited and exploratory studies of \glspl{ip} in \gls{qcd} \cite{Deryagin:1992rw, Muller:2013tya} suggest that \glspl{ip} or, in general, regimes with spatially oscillatory behavior are relevant in \gls{qcd} at finite density. 
	
	\Acrlongpl{ip} are very common in $\left(1+1\right)$-dimensional models at least within the mean-field approximation \cite{Thies:2003kk, Thies:2006ti, Thies:2018qgx, Thies:2019ejd, Thies:2020gfy, Thies:2020ofv, Koenigstein:2021llr, Thies:2023dyh}.
	However, in recent literature there is an on-going discussion whether \glspl{ip} persist when allowing for bosonic quantum fluctuations \cite{Lenz:2020bxk, Lenz:2021kzo, Lenz:2021vdz, Stoll:2021ori, Ciccone:2022zkg, Koenigstein:2023wso, Ciccone:2023pdk}.
	In $\left(3+1\right)$ dimensions, \glspl{ip} have been observed in various models over the last decades \cite{Kutschera:1990xk, Kutschera:1991rh, Nickel:2009wj, Carignano:2014jla, Buballa:2018hux, Carignano:2019ivp, Buballa:2020xaa}.
	However, recent studies \cite{Pannullo:2022eqh, Pannullo:2023ooo} show an inherit dependence of the \gls{ip} on the regulator value and the choice of the regularization scheme in the \gls{njl} model caused by the chemical potentials and momenta of the inhomogeneous condensates being in the order of the regulator.
	Thus, one can argue that there is no predictive power of the \gls{njl} model results with respect to \glspl{ip}\footnote{It is probably necessary to explore the regularization dependence of the \gls{ip} in models where the regulators can be tuned to higher values than in the \gls{njl} -- such as, e.g., the \gls{qm} model.}, as long as there is no good argument that an effective theory for \gls{qcd} at finite density could behave as a low-dimensional or strongly-regulated \gls{njl} model \cite{Pannullo:2023cat, Koenigstein:2023yzv}. 
	In contrast, the moat regime seems to be a stable feature of the \gls{njl} model \cite{Pannullo:2023ooo} -- as will be reported on in an upcoming publication following \Rcite{Pannullo:2023ooo}. 
	
	Due to the non-renormalizability of \gls{ff} models in $\left(3+1\right)$ dimensions and the problems arising from this non-renormalizability as well as due to the application to condensed matter systems, it is common to study these models in $\left(2+1\right)$ dimensions \cite{Klimenko:1987gi, Rosenstein:1988pt, Hands:1992ck, Inagaki:1994ec, DelDebbio:1995zc, Appelquist:2000mb, Hands:2003dh, Ebert:2015hva, Mandl:2022ryj, Lenz:2023wvk, Hands:2000gv}, where \gls{ff} models are renormalizable.
	In a preceding work \cite{Pannullo:2023one}, we have shown the absence of \glspl{ip} and moat regimes in a variety of \gls{ff} models and Yukawa models by analyzing the stability of homogeneous ground states.
	These results are in consistency with previous findings that minimize the effective action of the $\left(2+1\right)$-dimensional \gls{gn} model on the lattice \cite{Narayanan:2020uqt, Buballa:2020nsi, Pannullo:2021edr, Winstel:2022jkk} and generalizes this result to the whole class of models with Lorentz-scalar interaction channels. 
	
	In this work, we expand on the analysis in \Rcite{Pannullo:2023one} by including vector interactions $\sim \left(\bar{\psi} \gamma_\nu \psi \right)^2$.
	In contrast to nuclear matter, where the Walecka-model -- featuring interactions with vector mesons -- is commonly used, the inclusion of vector interactions is often not part of effective model calculations of \gls{qcd} matter except for a few studies \cite{Sasaki:2006ws, Redlich:2007rw, Carignano:2010ac, Carignano:2018hvn, Haensch:2023sig}.
	However, the fundamental \gls{qcd} action gives rise to Yukawa-type interactions between fermions and vector mesons, such as the $\omega$ meson, arising from a resonance of the \gls{ff} interaction in the corresponding vector channels \cite{Braun:2014ata, Rennecke:2015eba, Fukushima:2021ctq}.
	At non-vanishing $\mu$, the temporal component of the $\omega$ meson couples directly to the density $\langle \psi^\dagger \psi \rangle$ via the Yukawa interaction $\sim \psi^ \dagger \omega_0 \psi$.
	Thus, this interaction is expected to play an important role in finite density \gls{qcd}. 
	
	In the present work, we focus on the effects of mixing between scalar and vector condensates on the $\left(\mu, T\right)$ phase diagram by analyzing the stability of homogeneous ground states. 
	For this sake a \gls{ff} model including both scalar and vector interactions is defined in \cref{sec:model}.
	The renormalization of the \gls{ff} interaction and the methods for the computation of homogeneous condensates are briefly described in \cref{sec:methomcond} and \cref{sec:renorm}. 
	The stability analysis including the computation of the Hessian matrix in field space -- featuring mixing of different condensates -- is presented in \cref{sec:stab_analysis}.
	Symmetries of the Hessian matrix and their relation to finite density \gls{qcd} are discussed in \cref{sec:sym}.
	Further implications of the analysis regarding oscillatory correlations functions are presented in \cref{sec:compl_conj_Hessian}.
	In \cref{sec:homresults}, we present original results on the homogeneous phase diagram of the theory. 
	The appearance of oscillatory correlation functions in a quantum pion liquid regime as well as the absence of \glspl{ip} are discussed in \cref{sec:stab_sym}.
	In \cref{sec:impl}, we show that our results are generic for all models with local \gls{ff} interactions based on a similar argumentation as in \Rcite{Pannullo:2023one} before concluding in \cref{sec:concl}.
	
	\section{Fermionic model with scalar-vector mixing \label{sec:model}}
	In order to study mixing induced by the interplay of a repulsive vector and an attractive scalar interaction, we study the action
	\begin{align}
		\S_{\mathrm{mix}}[\bar{\psi},\psi]&   \label{eq:FFmodel} = \int_0^\beta \!  \dr \tau \int\! \dr^2 \xstV \Big\{ \bar{\psi}\left(\slashed{\partial}+ \gamma_3 \mu \right)  \psi - \left[ \tfrac{\coupling_S}{2 \N } \left(\bar{\psi}\,\psi\right)^2 + \tfrac{\coupling_V}{2 \N } \left( \left(\bar{\psi}\, \ii  \gamma_3 \, \psi\right)^2 + \left(\bar{\psi}\, \ii \vec \gamma \, \psi\right)^2 \right)  \right]\Big\} , 
	\end{align} 
	where $\psi$ contains $\N$ four-component spinors, describing a generic fermion field without a bare mass, $\mu$ is the chemical potential and $\gamma_\nu = \left(\gamma_3, \vec{\gamma}\right)$ are the $4\times4$ Dirac matrices\footnote{These fulfill the Clifford algebra with the Euclidean metric $\mathrm{diag}\left(1, 1, 1\right)$.}, where $\vec \gamma = (\gamma_1, \gamma_2)$.
	The spacetime integration goes over a $\left(2+1\right)$-dimensional Euclidean spacetime volume, where the inverse temperature determines the temporal extent $\beta = 1/ T$.
	Our conventions for the Wick rotation are given in \cref{app:wick}. 
	
	Introducing auxiliary bosonic fields in an inverse, shifted Gaussian integration yields the partially bosonized model
	\begin{align}
		\S[\bar{\psi}, \psi, \sigma, \omega_\nu] &= \int \mathrm{d}^3 x \, \left[\bar{\psi}\, \Q\, \psi + \frac{\omega_\nu \omega_\nu}{2\lambda_V} + \frac{\sigma^2 }{2\lambda_S} \right], \nonumber \\ \, Z &= \int \prod_{\phi=\{\bar{\psi}, \psi, \sigma, \omega_\nu\}} \mathcal{D} \phi \; \e^{-S[\bar{\psi}, \psi, \sigma, \omega_\nu]}  \label{eq:bos_action}
	\end{align}
	with the Dirac operator
	\begin{equation}
		\Q[\sigma, \omega_\nu] = \gamma_\nu \partial_\nu + \gamma_3 \mu + \sigma + \ii \gamma_\nu \omega_\nu. \label{eq:dir_op}
	\end{equation}
	The scalar field $\sigma$ and the vector field $\omega_\nu = (\omega_3, \vec{\omega})$ are linked to fermionic expectation values via
	\begin{align}
		\langle \sigma\rangle(x) &= - \tfrac{\coupling_S}{\N} \langle \bar{\psi}  \psi\rangle(x), \, \langle \omega_3 \rangle(x) = - \tfrac{\coupling_V}{\N} \ii \langle \bar{\psi}  \gamma_3 \psi\rangle(x), \, \langle \vec{\omega} \rangle(x) = - \tfrac{\coupling_V}{\N} \ii \langle \bar{\psi} \vec{\gamma} \psi \rangle(x). \label{eq:Ward}
	\end{align} 
	Assuming that the theories' invariance under rotations of the spatial coordinates $\mathbf{x} = \left(x_1, x_2\right) $ remains intact, only condensation of the scalar field $\sigma$ and the temporal component of the vector field $\omega_3$ is allowed. 
	Thus, this assumption implies that $\langle \vec{\omega} \rangle = 0$. 
	An inhomogeneous chiral condensate, however, can violate rotational invariance, which would invalidate the assumption of $\langle \vec{\omega} \rangle = 0$ . In this work, however, a stability analysis about homogeneous condensates is used, which respects the invariance of the theory under spatial rotations as it only depends on the absolute value of the spatial momentum vector $\mathbf{q}$ of the inhomogeneous perturbation by derivation, see section III of \Rcite{Buballa:2020nsi} and appendix B of \Rcite{Pannullo:2023one}. 
	No assumptions on the functional form of $\sigma(\mathbf{x})$ or $\omega_\nu(\mathbf{x})$ are made and the analysis only requires the insertion of homogeneous ground states as expansion points. 
	Thereby, rotational invariance is respected by this analysis such that the assumption of $\langle \vec{\omega} \rangle = 0$ remains valid whenever information about the thermodynamic ground state of the system is obtained.
	A non-vanishing expectation value of $\sigma$ indicates the breaking of the discrete chiral symmetry of this model \cite{Klimenko:1987gi, Buballa:2020nsi, Pannullo:2021edr}, since it acts as a dynamically generated mass term for the fermions. 
	This mass term is parity-even, while there is also the possibility for a parity-odd mass term in $\left(2+1\right)$ dimensions \cite{Pisarski:1984dj, Pannullo:2023one}.
	Thus, we often refer to $\langle \sigma \rangle$ as the chiral condensate, since it is directly linked through \cref{eq:Ward} to $\langle \bar \psi \psi\rangle$.
	Also, we note that in our conventions (compare \cref{app:wick}) the quark number density is given by $n(x) =-\langle \bar{\psi} \gamma_3 \psi \rangle (x) / \N$ and, thus, $\omega_3$ is required to be a purely imaginary field with $\Im\omega_3(x) = \lambda_V n(x) / \N$ in order for the baryon density to be real-valued. This is directly linked to the repulsive nature of the Yukawa interaction between quarks and the vector meson $\omega_3$ and the corresponding \gls{ff} interaction.
	
	\subsection{Mean-field approximation}
	
	For the remainder of this work, we use the so-called mean-field approximation, which in this context means the suppression of bosonic quantum fluctuations in the partition function \eqref{eq:bos_action}.
	In the case of \gls{ff} models such as \cref{eq:FFmodel}, this suppression can directly be obtained by taking the limit of the parameter $\N \rightarrow \infty$. 
	Since the fermion fields appear only as bilinears in \cref{eq:bos_action}, one can integrate their fluctuations out. 
	The obtained effective action\footnote{In the mean-field approximation, the effective action is directly proportional to the quantum effective action, usually defined as the Legendre transform with respect to of the logarithm of the partition function.}
	\begin{align}
		\frac{1}{\N} \seff[\sigma, \omega_\nu] = \int \mathrm{d}^3 x \, \left[ \frac{\omega_\nu \omega_\nu }{2\lambda_V} + \frac{\sigma^2}{2\lambda_S}  \right] - \Tr \ln \Q \label{eq:effaction}
	\end{align}
	is then a functional of the bosonic fields only. 
	In the mean-field approximation, observables can be computed by evaluating them on the global minimum of the effective action\footnote{In the case of multiple, degenerate global minima, which are linked through a symmetry transformation, one has to formally introduce a small symmetry breaking parameter $h$ and extrapolate to a vanishing $h$. In the mean-field approximation, this can be directly implemented by picking one of the degenerate minima. In the case of obtaining a point in the phase diagram with a first order phase transition, where this issue can lead to ambiguities, we will refrain from evaluating any observables depending on the minimum of the effective action. We refer to \Rcite{Pannullo:2023one} for a similar discussion.}, denoted by $\sigma = \minHs$ and $\omega_\nu = \Omega_\nu$. 
	Since the interactions do not mix between the $N$ fermion fields, the effective action is proportional to $\N$ and the minimization of $\seff$ is independent of $\N$. 
	Hence, the obtained phase diagram would be identical when using the mean-field method as a semi-classical approximation of the \gls{qft} with any integer number of $\N$. 
	The mean-field approximation has turned out to be a decent starting point for investigations of phase diagrams, since ordered phases as, e.g., an \gls{hbp} or an \gls{ip}, are typically weakened through bosonic quantum fluctuations, see, e.g., \Rcite{Lenz:2020bxk, Lenz:2021kzo, Lenz:2021vdz, Lenz:2023wvk,Stoll:2021ori, Ciccone:2022zkg}.
	Thus, if an \gls{hbp} or an \gls{ip} does not exist in the mean-field approximation, it is very likely that those phases also do not exist in the full \gls{qft}.
	
	\section{Methods for computing the phase diagram in the presence of mixing \label{sec:methods}}
	In this chapter, we present methods for computing the phase diagram of the \gls{ff} model with mixing, see \cref{eq:FFmodel}. 
	
	\subsection{Homogeneous condensates \label{sec:methomcond}}
	As a starting point for the investigation of the phase structure of the model \eqref{eq:FFmodel}, the thermodynamic ground state with the assumption of homogeneous condensates $\sigma = \Hom{\sigma},\,\omega_3 = \Hom{\omega}_3$ is determined. 
	We assume that the spatial components of the vector field respect the invariance of the partition function \eqref{eq:bos_action} under spatial rotations, i.e. $\vec{\Hom{\omega}} = 0$. 
	With this assumption one can compute the homogeneous effective potential at fixed $\mu$ and $T$ 
	\begin{equation}
		\HomPot^{(\mu,T)}(\Hom{\sigma}, \Hom{\omega}_3) = \frac{1}{\N} \frac{\seff[\Hom{\sigma}, \Hom{\omega_3}]}{\beta V}
	\end{equation}
	using standard techniques of thermal field theory.
	Inspecting \cref{eq:bos_action} under the assumption of $\omega_3 = \Hom{\omega}_3$, it is convenient to absorb the vector condensate into an effective chemical potential $\muSh = \mu + \ii \Hom{\omega} = \mu - \lambda_V \Hn $, where $\Hn$ is the homogeneous quark number density per fermion species, see the discussion below the Ward identity \eqref{eq:Ward}. 
	The homogeneous effective potential is given by 
	\begin{align}
		\HomPot^{(\mu,T)}(\Hom{\sigma}, \Hom{\omega}_3) &=  \frac{\Hw^2}{2 \lambda_V} + \frac{\Hs^2}{2 \lambda_S} - \frac{N_\gamma}{2} \int \frac{\dr^2 p}{\left(2 \pi\right)^2} \left\{ E + \frac{1}{\beta} \Bigg[\ln\left(1 + \e^{-\beta(E + \muSh)}\right) + \muSh \rightarrow - \muSh \Bigg]\right\},\label{eq:hompot}
	\end{align}
	where the last two terms are identical to the effective potential of the \gls{gn} model at chemical potential $\muSh$ and temperature $T$, as, e.g., studied in \Rcite{Klimenko:1987gi, Rosenstein:1988pt, Narayanan:2020uqt, Buballa:2020nsi, Koenigstein:2021llr}. 
	In \cref{eq:hompot}, summation over fermionic Matsubara frequencies was already performed and $E = \sqrt{\vec{p}^2 + \Hom{\sigma}^2}$, where $\vec{p}$ are the spatial momenta. 
	The dimensionality of the spinors is denoted by $N_\gamma = 4$.  
	
	The computation of homogeneous condensates is standard in the literature, see, e.g., \Rcite{Klimenko:1987gi, Thies:2003kk, Buballa:2003qv, Buballa:2014tba, Koenigstein:2021llr}. In order to determine the extrema of the effective potential \eqref{eq:hompot} with respect to $\Hs$ and $\Hw$, we use the so-called gap equations
	\begin{align}
		\frac{\partial \HomPot^{(\mu,T)}}{\partial \bar \sigma} &= 0 \leftrightarrow  \Hs \left(\frac{1}{\lambda_S} - \ell_1(\mu,T, \Hs, \Hw) \right) = 0, \label{eq:GapSigma}
	\end{align}
	where
	\begin{equation}
		\ell_1(\mu,T, \Hs, \Hw) = \frac{N_\gamma }{2} \int \frac{\dr^2 p}{\left(2\pi\right)^2} \frac{1-\FD(E) - \AFD(E) }{E},  
	\end{equation}
	and  
	\begin{align}
		\frac{\partial \HomPot^{(\mu,T)}}{\partial \bar \omega} = 0 \leftrightarrow \frac{\Hw}{\lambda_V} - \ii \Hn(\mu, T, \Hs, \Hw) = 0 \label{eq:GapOmega}
	\end{align}
	with 
	\begin{equation}
		\Hn(\mu, T, \Hs, \Hw) =  \frac{N_\gamma}{2} \int \frac{\dr^2 p}{\left(2\pi\right)^2} \left[\FD(E) - \AFD(E)\right].
	\end{equation}
	In the above expressions, $\FD(x) = \left(1 + \e^{\beta (x-\muSh)}\right)^{-1}$ and $\AFD(x) = \left(1 + \e^{\beta (x+\muSh)}\right)^{-1}$. 
	Note that the gap equation for $\Hw$ does not contain a vacuum contribution from the Dirac see by derivation, such that $\Hw = 0$ is the only solution of \cref{eq:GapOmega} for $\mu = 0$.
	Solving this set of coupled equations, one computes all extrema of $\HomPot$ with respect to $\Hs$ and $\Hw$ and determines the global minimum, denoted as $\left(\minHs(\mu, T), \minHw(\mu,T)\right)$, respectively, by inserting the all found extrema back into the effective potential \eqref{eq:hompot}. 
	
	The absorption of $\bar{\omega}_3$ into a chemical potential $\muSh$ can be useful for the interpretation of results as well as for computational purposes. 
	As there is no way to distinct the dynamical contribution to the chemical potential through the condensation of the vector field $\minHw$ experimentally, $\muSh$ is also the phenomenologically relevant quantity. 
	Thus, one could argue to use $\muSh$ and $T$ as external parameters of the model and, instead, treat  the effective potential as a function of $\Hs$ and $\mu$, such as done in \Rcite{Sasaki:2006ws, Redlich:2007rw}. 
	This is also advantageous for the minimization of the effective potential \cref{eq:hompot}, since one can directly solve \cref{eq:GapSigma} at fixed $\muSh$ and $T$.
	Inserting the solution $\minHs$ into \cref{eq:GapOmega}, it is possible to directly compute $\minHw$ and, thus, the chemical potential from $\mu = \muSh - \ii \minHw$. 
	However, this procedure leads to ambiguities in the $(\mu, T)$ phase diagram, as multiple $(\Sigma, \muSh)$ pairs can belong to the same chemical potential $\mu$.
	Then, one would need to iterate through all computed tuples $\left(\sigma, \muSh\right)$ to get the correct global minimum of $\HomPot$ for a certain $(\mu, T)$ and draw a correct picture of the phase diagram.
	Thus, in our numerical setup we keep $\Hs$ and $\Hw$ as variables.
	
	\subsection{Renormalization and parameter fixing \label{sec:renorm}}
	As can be seen from \cref{eq:GapSigma}, the computation of homogeneous condensates in the model \eqref{eq:FFmodel} involves the evaluation of integrals with linear \gls{uv} divergences that require regularization. 
	As originally demonstrated in \Rcite{Rosenstein:1988pt, Gat:1991bf}, \gls{ff} interactions are renormalizable order by order in $1/\N$ in $\left(2+1\right)$ dimensions. 
	Thus, we can remove the divergences by imposing a renormalization condition. 
	By fixing the dynamically generated fermion mass in the vacuum $\minHs(\mu=0, T=0) = \so$, one can absorb the divergence using the coupling $\lambda_S$ through the gap equation \eqref{eq:GapSigma} in the vacuum.  
	Using a sharp \gls{uv} cutoff $\Lambda$ for the spatial loop momenta, we determine
	\begin{equation}
		\ell_1(0, 0, \Hs, 0) = \frac{\N_\gamma}{4\pi}\left(\Lambda - |\Hs|\right)
	\end{equation} 
	such that
	\begin{equation}
		\frac{1}{\lambda_S} = \frac{\N_\gamma}{4\pi}\left(\Lambda - |\so|\right).
	\end{equation}
	Note that we used that $\Hw = 0$ in the vacuum, as can be read off directly from \cref{eq:GapOmega}. Using this procedure and sending $\Lambda \rightarrow \infty$, we obtain
	\begin{align}
		L_1&(\mu, T, \Hs, \Hw) \equiv \frac{1}{\lambda_S} - \ell_1(\mu, T, \Hs, \Hw) = \N_\gamma \left[\frac{|\Hs| - |\so|}{4 \pi} + \int \frac{\dr^2 p}{(2\pi)^2} \frac{\FD(E) + \AFD(E)}{2 E}\right],
	\end{align}
	which is finite and is all that is required to render the observables of interest in this work finite. 
	Thus, the chiral condensate in the vacuum $\so$ will be used as the physical scale to construct dimensionless ratios for all other quantities in this work. 
	
	The gap equation for $\Hw$ \eqref{eq:GapOmega} does not contain a divergent vacuum contribution, such that imposing a renormalization prescription in the common way is not possible\footnote{Also, to the knowledge of the author there is no alternative way of renormalizing this coupling in the literature.}.
	Moreover, the \cref{eq:GapOmega} yields $\Hw = 0$ as the only solution in the vacuum.
	Thus, $\lambda_V$ can only be fixed by working at $\mu \neq 0$, e.g., by imposing a value for $\Hw$ at a non-vanishing $\mu$ and $T = 0$. 
	Since we consider the action \eqref{eq:FFmodel} as a toy model for the mixing of scalar and vector modes, we do not fix $\lambda_V$ but treat it as a free parameter to study its influence on the phase diagram of the theory similar to \Rcite{Sasaki:2006ws, Redlich:2007rw}.

	\subsection{The Hessian matrix and bosonic two-point vertex functions in the presence of mixing \label{sec:stab_analysis}}
	In order to compute the phase structure of the theory, the ground state $\left(\sigma, \omega_\nu\right) = \left(\Sigma(\mathbf{x}), \Omega_\nu(\mathbf{x})\right)$ needs to be determined. 
	Without specifying a particular ansatz for the dependence of the condensates on the spatial coordinates, this is a difficult functional minimization problem, that was so far not consistently solved in the literature.
	However, an efficient strategy to test whether, e.g., an \gls{ip} is favored compared to homogeneous ground states $\vminauxf_j$ is to perform a stability analysis, where the homogeneous condensates are perturbed by arbitrary functions of infinitesimal amplitude $\ptauxf_j(\mathbf{x})$. 
	References \cite{Koenigstein:2021llr, Koenigstein:2023wso} give an in-depth discussion of the method and its advantages and drawbacks on the example of a model with an analytical solution for $\Sigma(\mathbf{x})$ in the whole $\left(\mu, T\right)$ plane.
	For a $2+1$-dimensional model, the same method was used in \Rcite{Buballa:2020nsi, Pannullo:2023one}.
	Additional implications of the analysis, that are not discussed in \Rcite{Koenigstein:2021llr, Koenigstein:2023wso}, will be subject of \cref{sec:compl_conj_Hessian}. 
	
	The main idea of the method is to determine whether the non-vanishing leading-order coefficients in an expansion with respect to the Fourier mode $\ptftauxf_j(\mathbf{q})$ of an inhomogeneous perturbation is negative and, thus, leads to a lower free energy.
	Thereby, $\mathbf{q}$ denotes the spatial momentum vector. 
	In this case there is an instability of homogeneous condensates $\Hminauxf_j$ (or linear combinations of the condensates) towards an \gls{ip}.
	The first non-vanishing contribution when perturbing around solutions of the gap Eqs.~\eqref{eq:GapSigma} and \eqref{eq:GapOmega} is of second order in the bosonic fields, since the first order contribution is proportional to the gap Eqs. \cite{Koenigstein:2021llr}.
	In the case of multiple bosonic fields, the second order contribution is given by a Hessian matrix $H_{\auxf_j \auxf_k}$, which needs to be diagonalized in field space by finding its eigenvalues $\Gamma^{(2)}_{\dbf_j}$ and eigenvectors $\dbf_{j}$, where, in general, $\dbf_{j} \neq \auxf_j$.
	In the case at hand, the eigenvalues as well as the elements of the Hessian matrix itself do not depend on the direction of $\mathbf{q}$, but only on its absolute value $q = |\mathbf{q}|$ as first derived in \Rcite{Buballa:2020nsi} for a $2+1$-dimensional \gls{ff} model. 
	This is caused by the locality of the fermion self energy stemming from incorporating only local interactions in \cref{eq:FFmodel}, as discussed in Sec.~IV of \Rcite{Motta:2023pks}.
	In practice, one obtains this result when performing the Fourier transformation of the second order correction as described in section III of \Rcite{Buballa:2020nsi}.
	In \Rcite{Pannullo:2023one} multiple examples of the diagonalization procedure are given.
	The eigenvalues $\Gamma^{(2)}_{\dbf_j}$ of the Hessian matrix are called the bosonic two-point vertex functions\footnote{In the mean-field approximation, this quantity is precisely the two-point-one-particle-irreducible vertex function, as it is often defined in \gls{qft} textbooks.} and are typically expressed as functions of the bosonic momentum $q$ corresponding to the respective Fourier mode of $\ptdbf_j$\footnote{The two-point function does not depend on the direction of the vector $\vec{q}$, because the computation involves the integration over loop-momenta, which can always be rotated such that the resulting integrals only depend on the absolute value of $q$ respecting rotational invariance.}.
	The stability analysis is equivalent to a saddle point approximation of the partition function \eqref{eq:bos_action} with respect to the bosonic fields.
	Investigations of this type have recently been used a lot in the literature, see, e.g., \Rcite{Nakano:2004cd, Pannullo:2021edr, Carignano:2019ivp, Buballa:2020nsi, Pannullo:2023one, Pannullo:2023cat, Koenigstein:2023yzv}. 
	
	For a detailed derivation of the bosonic two-point vertex functions in models with multiple \gls{ff} or Yukawa-type interactions we refer to appendix A of \Rcite{Pannullo:2023one}.
	The focus of this Sec.~is on the important differences in the analysis compared to the one in \Rcite{Pannullo:2023one} caused by the included vector interactions. 
	As discussed in the aforementioned Refs., the Hessian matrix can be derived by considering all second-order functional derivative terms with respect to the bosonic fields $\auxf_j$.
	One then has to find a basis transformation mapping $\auxf_j$ on a field basis $\dbf_j$ diagonalizing $H$ such that the eigenvalues of $H$ are the two-point vertex functions $\Gamma^{(2)}_{\dbf_j}(q)$.
	With Lorentz-scalar interaction, one usually finds $\dbf_j = \auxf_j$.
	This is not the case, when studying a model with vector interactions, such as \cref{eq:bos_action} with $\vauxf = \left(\sigma, \omega_\nu\right)$. 
	Then, not only $\dbf_j \neq \auxf_j$ but also the eigenvectors of $H$ in field space have to be determined for every value of $q$, i.e.~$ \dbf_j(q) = \sum_k c_{j,k}(q) \auxf_k$. 
	
	In a model with a repulsive vector interaction, as mediated by $\omega_3$, the situation gets even more complicated. As discussed above, the homogeneous condensate $\Hw \sim \ii \Hn$ is purely imaginary. 
	In contrast to the real-valued condensate $\Hs$ being analyzed by studying its stability with respect to real valued perturbations $\delta \sigma$, one has to treat fluctuations $\delta \omega_3$ about $\Hw$ in the complex plane.
	As recently discussed in \Rcite{Haensch:2023sig}, fluctuations must be in the direction of the steepest descent of the effective action corresponding to the stable Lefshetz thimble. 
	The analysis of fluctuations about the homogeneous ground state is equivalent to saddle point approximation of the path integral, which in turn is only well-defined when the stable Lefshetz thimble is used \cite{Alexandru:2020wrj}. 
	Inspecting the effective action \cref{eq:bos_action} under these aspects reveals that one has to consider real-valued fluctuations $\delta \omega_3$ about the purely imaginary $\Hw$.
	This is demonstrated in a similar, \gls{njl}-type model with vector interactions in $\left(3+1\right)$ dimensions in \Rcite{Mori:2017zyl}.
	In Figure 5 therein, the Lefshetz thimble in the field space of the temporal vector component is depicted. \\

	The resulting Hessian for model \eqref{eq:bos_action} is given by
	\begin{widetext}
		\begin{align}
			H_{\auxf_j \auxf_k}(q) = \delta_{\auxf_j,\auxf_k} \left(\frac{\delta_{\auxf_j,\sigma}}{\lambda_S} + \frac{1-\delta_{\auxf_j,\sigma}}{\lambda_V}  \right) + \frac{1}{\beta} \sum_n \int \frac{\dr^2 p}{(2\pi)^2} \tr\left[\Prop(\nu_n, \vec{p}+ \vec{q})\, c_j \, \Prop(\nu_n, \vec{p})\, c_k\right], \label{eq:Hessian}
		\end{align}
	\end{widetext}
	where we define the propagator of a free fermion with mass $\minHs$ at chemical potential $\muSh = \mu + \ii \minHw$ and temperature $T$ as
	\begin{equation}
		S = \left(\hat{\Q}[\minHs, \minHw\delta_{\nu, 3}]\right)^{-1},
	\end{equation}
	where $\hat{\Q}$ denotes the Fourier transform of $\Q[\sigma, \omega_\nu]$, compare \cref{eq:dir_op}. 
	Also, $\nu_n =  2\pi(n - \frac{1}{2})T$ are fermionic Matsubara frequencies, $\vec{q} = \left(q,0\right)$, where the spatial momentum integration can be chosen such that $q$ is aligned along one coordinate axis and we defined the bare vertices $\vec{c} = \left(1, -\gamma_3, \ii \gamma_1, \ii\gamma_2\right)^T$.
	Note that purely imaginary perturbations about $\Hw$ would correspond to a vertex $\ii \gamma_3$ instead of $-\gamma_3$.
	From \cref{eq:Hessian}, the cyclic property of the trace gives $H_{\auxf_j \auxf_k} = H_{\auxf_k \auxf_j}$.
	Formulas for the evaluation of $H_{\auxf_j \auxf_k}(q)$ can be found in \cref{app:stab_analysis}.
	We note that \cref{eq:Hessian} describes a fermionic one-loop diagram with amputated bosonic legs (compare Eq.~(36) in \Rcite{Koenigstein:2021llr}) already incorporating mixing effects between the bosonic fields in the mean-field approximation.
	One finds $H_{\sigma \omega_\nu} \propto \minHs$, such  mixing between the scalar and the vector mode is only relevant within the \gls{hbp}. 
	In practical computations, we compute $H_{\auxf_j \auxf_k}(q)$ for a fixed $q$ and diagonalize the resulting matrix numerically using \textit{Python3} with various libraries \cite{Python3, Virtanen:2019joe, Harris_2020}.

	\subsection{Symmetries of the Hessian \label{sec:sym}}
	In the following, we will analyze the structure of the Hessian \eqref{eq:Hessian}. 
	A similar discussion can be found in the recent work \cite{Haensch:2023sig}, where a Polyakov-loop \gls{qm} model with vector mesons has been studied using a static stability analysis, i.e., the Hessian was studied at $q=0$ only.
	Since $\omega_\nu$ are components of a vector field, they pick up a sign under charge conjugation, i.e., $\mathcal{C} \omega_\nu \mathcal{C}^{-1} = - \omega_\nu$ (see \Rcite{Scherer:2012nn} for a construction of the charge conjugation operation on four-dimensional spinors in $\left(2+1\right)$ Euclidean spacetime dimensions).
	This breaking of charge conjugation symmetry is expected at finite density due to the excess of particles over anti-particles.
	Since $H_{\sigma \omega_3} =  H_{\omega_3 \sigma}$ is non-vanishing at $\mu \neq 0$ and purely imaginary, we conclude that the Hessian is non-Hermitian when $\mathcal{C}$-symmetry is broken. 
	On the level of homogeneous condensation the breaking of $\mathcal{C}$-symmetry is realized through $\Hw \neq 0$ at $\mu \neq 0$, as will be discussed in \cref{sec:homresults}.
	However, the homogeneous condensates as well as the Hessian are still invariant under the combined operation of the linear $\mathcal C$ transformation and anti-linear transformation of complex conjugation $\mathcal K$.
	A remaining invariance under a combined anti-linear $\mathcal C \mathcal{K}$ operation while the charge conjugation symmetry is broken at finite density can also be found in \gls{qcd} itself as well as other \gls{qcd}-inspired theories, such as Polyakov-\gls{njl} models \cite{Nishimura:2014kla, Felski:2020vrm, Haensch:2023sig}. 
	
	Due to the invariance under the $\mathcal C \mathcal{K}$ operation the Hessian obeys the relation $H = A H^* A$, where $^*$ denotes complex conjugation and $A= \mathrm{diag}(1, -1,1,1)$. 
	It directly follows that $H(q)$ possesses the same set of eigenvalues ${\lambda_j}$ as $H^*(q)$ and, consequently, the eigenvalues are either real-valued or come in complex-conjugate pairs. 
	In the case of real-valued $\lambda_j(q)$, their interpretation as bosonic two-point vertex functions $\Gamma^{(2)}_{\dbf_{j}}(q)$ has been extensively discussed in the previous section and also in \Rcite{Pannullo:2023one}.
	However, it has recently been demonstrated that the competition between repulsive and attractive interactions can induce complex-conjugate eigenvalue pairs in the stability analysis based on the study of mixing in Euclidean field theories with $\mathcal{P} \mathcal{T}$-type symmetries\footnote{Here, $\mathcal{P}\mathcal{T}$-type symmetry means symmetry under an linear symmetry operation $\mathcal{P}$ and an arbitrary anti-linear symmetry operation $\mathcal{T}$. Field theories with $\mathcal{P} \mathcal{T}$-type symmetries typically describe theories, that are not invariant under either $\mathcal{P}$ or $\mathcal{T}$ separately, but show invariance under their combined operation. $\mathcal{P} \mathcal{T}$-type symmetries are also widely studied in quantum mechanics, optics and condensed matter for their unique properties \cite{Bender:1998ke, El-Ganainy:2018ksn, MiriAlu:2019kn, Ashida:2020dkc, Bender:2023cem}. Also, they are also known to be useful in resolving sign problems in lattice field theory simulations, see, e.g., \Rcite{Meisinger:2012va, Ogilvie:2018fov, Schindler:2019ugo, Ogilvie:2021wvb}.} \cite{Schindler:2019ugo, Schindler:2021otf, Schindler:2021cke}.
	The existence of complex-conjugate eigenvalue pairs leads to bosonic correlation functions with spatial sine-like modulations in addition to the ordinary exponential decay and is directly related to the invariance under a combination of an linear and an anti-linear symmetry operation, such as the $\mathcal C \mathcal{K}$ operation \cite{Nishimura:2014kla}.

	\subsection{Accessing properties of bosonic two-point correlation functions within the mean-field approximation using the Hessian matrix \label{sec:compl_conj_Hessian}}
	
	The arguments for the existence of these oscillating bosonic correlation functions in the literature and the validity of the analysis in the case of the present model are briefly recapitulated here.
	
	\subsubsection{Recapitulation of the literature on Hessian matrix analysis in bosonic field theories}
	As discussed in \Rcite{Schindler:2019ugo} and \Rcite{Schindler:2021otf}, the inverse propagator matrix of dynamical bosonic fields is given by 
	\begin{equation}
		q^2 + H(0)
	\end{equation}
	where the $q^2$-term comes from an a priori kinetic term as included in \Rcite{Schindler:2019ugo, Schindler:2021cke, Schindler:2021otf} and $H(0)$ is the static Hessian matrix (equivalent to the Mass matrix in the discussed theories in these references).
	This expression stems from the fact that the inverse of the bosonic correlation functions can be expressed as a sum of the tree level contributions and self-energy contributions stemming from the bosonic one-loop analysis. 
	In this case, one can directly classify the behavior of propagators by just inspecting the eigenvalues $\Gamma^{(2)}_{\dbf_j}(0)$ of the static Hessian $H(0)$.
	The classifications stem from the fact that roots of the expression $q^2 + H(0)$ correspond to poles of the corresponding propagators of the dynamical bosonic fields in the one-loop analysis.
	Ordinary, stable homogeneous phases with exponentially decaying propagators yield Hessian matrices with positive eigenvalues.
	An even number of negative eigenvalues corresponds to a stable ground state with respect to homogeneous perturbations but instable with respect to inhomogeneous ones.
	Based on this analysis, one can obtain indications about \glspl{ip} without including the full-momentum dependence of $H$.
	For an uneven number of negative eigenvalues in the static Hessian, the system is completely unstable against both homogeneous and inhomogeneous perturbations corresponding to a set of field values, which are not a stable minimum of the system. 
	The complex-conjugate eigenvalues give rise to what is defined in \Rcite{Schindler:2021cke} as modulated exponential decay, where one finds spatial oscillations with a momentum scale related to the imaginary part of the eigenvalue of $H(0)$. 
	The real part of the eigenvalue is then a scale for the exponential decay, as it is with ordinary screening poles.
	This behavior \cite{Akerlund:2016myr} is also observed when an inhomogeneous ground condensate, such as, e.g., the chiral spiral, is destabilized through the fluctuations of Goldstone modes of $\mathrm O(N)$ symmetry breaking \cite{Pisarski:2020dnx}.
	This analysis can be seen as an extension of the common stability analysis discussed before.
	However, it has, to our knowledge, so far only been performed with static modes \cite{Rennecke:2023xhc}. 
	
	\subsubsection{Implications for the analysis of bosonic two-point correlation functions within the mean-field approximation in the four-fermion model}
	In contrast to the previous works \cite{Rennecke:2023xhc,Schindler:2021cke}, we consider auxiliary fields and, thus, do not a priori include a kinetic term for the bosonic fields.
	Then, the bosonic two-point vertex functions $\Gamma^{(2)}_{\dbf_j}(q)$ are the inverse of the bosonic two-point correlation functions $G_{\dbf_{j}}$ for the fields $\dbf_j$.
	This can be obtained from studying the Dyson equation for the two-point correlation functions, recalling that the bosonic self-energy in the Dyson equation is given by the fermion-loop contribution in the mean-field approximation. 
	Then, the inverse of the two-point correlation functions are given by the so-called two-point (one-particle-irreducible) vertex function given by the second functional derivative of the quantum effective action with respect to the bosonic fields evaluated at their expectation values, see, e.g., \Rcite{Rivers1987}. 
	In the mean-field approach, the quantum effective action is approximated by the effective action \eqref{eq:effaction} evaluated at its minimum and, thus, the two-point vertex functions $\Gamma^{(2)}_{\dbf_j}(q)$ in turn are equivalent to the eigenvalues of the Hessian matrix \eqref{eq:Hessian}, which is proportional to the second derivative of the effective action \cref{eq:effaction} with respect to the bosonic fields. 
	The low-momentum expansion of the bosonic two-point vertex functions results in
	\begin{equation}
		G_{\dbf_{j}}^{-1} = \Gamma^{(2)}_{\dbf_j}  \approx Z_{\dbf_j} q^2 + \Gamma^{(2)}_{\dbf_j}(0) \label{eq:invprop_approx}
	\end{equation}
	generating a kinetic term in the quantum effective action, where we defined the bosonic wave-function renormalization 
	\begin{equation}
		Z_{\dbf_j} = \frac{1}{2} \frac{\dr^2 \Gamma^{(2)}_{\dbf_j}}{\dr q^2}\Big|_{q=0} \label{eq:wavefunc_gen}
	\end{equation}
	as the coefficient of the kinetic term generated by the fermion loop. 
	Thus, one can perform an analysis similar to the case of bosonic theories described by finding the roots of \cref{eq:invprop_approx}.
	It is important to note that $Z_{\dbf_{j}}$ can have a small\footnote{By small, it is meant that the imaginary part is typically at least two orders of magnitude smaller than the real part.}, but non-vanishing imaginary part.
	As one can see from \cref{eq:invprop_approx}, both $\Gamma^{(2)}_{\dbf_{j}}(0)$ and $Z_{\dbf_{j}}$ being real-valued leads to purely imaginary propagator poles such that one obtains the usual exponential decay of $G_{\dbf_{j}}$.
	However, a non-vanishing real part of the obtained roots yields the above described oscillatory behavior of the propagator.

	The analysis of the momentum-dependent Hessian $H(q)$ is, however, more elaborate.
	In \cref{sec:results} we will obtain the existence of regimes with complex-conjugate bosonic two-point vertex functions $\Gamma^{(2)}_{\dbf_j}(q)$ at both $q = 0$ and $q\neq 0$, but also regimes in the $\left(\mu,T\right)$ phase diagram of the \gls{ff} model \eqref{eq:bos_action}, where $\Gamma^{(2)}_{\dbf_j}(q=0) \in \mathbb{R}$ but $\Gamma^{(2)}_{\dbf_j}(q\neq 0) \in \mathbb{C}$.
	To our knowledge, the interpretation of $H(q)$ developing complex-conjugate eigenvalue pairs at $q \neq 0 $ while the eigenvalues at $q=0$ are real-valued is unclear.
	Especially, the low-momentum expansion of the two-point correlation function is not meaningful at non-vanishing momenta as one computes the Hessian by expanding about the homogeneous field values $\left(\sigma, \omega_\nu\right) = \left(\minHs,\, \minHw \delta_{\nu, 3}\right)$.
	Statements with respect to the existence of an \glspl{ip} can, however, be made as discussed above, as long as the bosonic two-point vertex functions are real-valued.
	
	\section{Phase diagram in the presence of mixing\label{sec:results}}
	We proceed by presenting our results for the phase diagram. 
	Thereby, the vector coupling $\lambda_V$ is treated as a free parameter and is varied to study varying strengths of the vector coupling.
	In order to define a scale for the strength of the scalar coupling $\lambda^R_S$ used for comparison to the used values of the vector coupling $\lambda_V$, we define $\lambda^R_S = 1 / \Gamma^{(2)}_\sigma(q = 0)$ in the vacuum. 
	Since the Hessian \cref{eq:Hessian} is diagonal in the vacuum, this yields $\lambda^R_S = \pi $, where the result for $\Gamma^{(2)}_\sigma$ in the vacuum can be found by computing \cref{eq:H_ss} in the limit of zero temperature and chemical potential, see Appendix A of \Rcite{Pannullo:2023cat} for the explicit expression.
	Thus, we consider $\lambda_V \so \in [0, \pi]$.
	Note that all dimensionful quantities in plots are expressed in units of the chiral condensate in the vacuum $\so$.

	\subsection{Homogeneous condensation \label{sec:homresults}}
	As the homogeneous ground states are the input for the stability analysis, we first compute the phase diagram of model \eqref{eq:FFmodel} when assuming homogeneous condensation $\left(\sigma, \omega_\nu\right) = \left(\Hs, \Hw\, \delta_{\nu, 3}\right)$.
	Minimizing the effective potential \cref{eq:hompot} with respect to $\Hs$ and $\Hw$ as described in \cref{sec:methomcond}, we determine the thermodynamic ground state $(\Hs, \Hw) = \left(\minHs, \minHw \right)$.
	We distinct between the \gls{sp}, where $\minHs = 0$ and chiral symmetry is restored, and the \gls{hbp}, where chiral symmetry is spontaneously broken by a non-vanishing $\minHs$.
	By studying the gap \cref{eq:GapSigma} and the effective potential \eqref{eq:hompot} in \cref{sec:methomcond}, one obtains that the phase boundary between \gls{hbp} and \gls{sp} in the $\left(\muSh, T\right)$ of the \gls{ff} model defined in \cref{eq:FFmodel} is identical to the one of the $\left(2+1\right)$-dimensional \gls{gn} model in the $\left(\mu,T\right)$ plane, as first determined in \Rcite{Klimenko:1987gi}.
	The analytic expression for the critical chemical potential $\mu_c$ of the \gls{gn} model is given by
	\begin{equation}
		\mu_c(T) = T \mathrm{arcosh}\left(0.5 \e^{\so/T} - 1\right). \label{eq:GNmuc}
	\end{equation}
	We use this analytical result as a cross-check for our numerical results at $\lambda_V \neq 0$, by computing $\muSh_c = \mu_c + \ii \minHw$ at the obtained phase boundary $\mu_c(\lambda_V, T)$ and comparing the result to the case without vector coupling $\mu_c(0.0, T)$, as given by the right hand side of \cref{eq:GNmuc}. 
	The phase boundary in the $\left(2+1\right)$-dimensional \gls{gn} model is of second order, except for the point $\left(\mu / \so=1.0, T / \so  = 0.0\right)$ where one obtains an effective potential, which is flat for $\Hs \in [0.0, 1.0]$. 
	
\begin{figure}[t]
		\centering
		\includegraphics{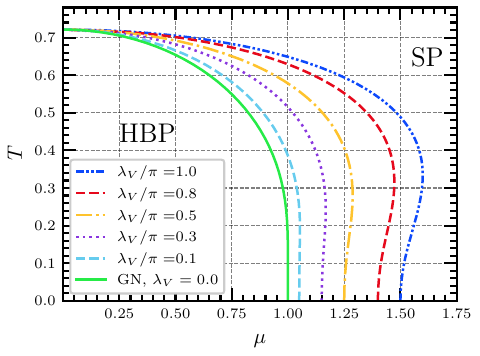}	
		\caption{\label{fig:homboundary} Phase boundary lines between the \gls{hbp} and the \gls{sp} in the $(\mu, T)$ space for 5 different values of the vector coupling $\lambda_V$. The $\lambda_V = 0.0$ phase boundary corresponds to the known analytic solution of the \gls{gn} model in $\left(2+1\right)$ dimensions \cite{Klimenko:1987gi, Buballa:2020nsi}.}
	\end{figure}

	In \cref{fig:homboundary}, we plot the phase boundary lines of the model \eqref{eq:FFmodel} for different values of $\lambda_V \so \in [0.0, \pi
	]$.
	Similar to findings with \gls{njl}-type models featuring vector interactions \cite{Sasaki:2006ws,Carignano:2010ac, Carignano:2018hvn}, an enlargement of the \gls{hbp} is observed when increasing the vector coupling.
	The extent of the \gls{hbp} in the $(\mu, T)$ plane grows monotonically with $\lambda_V$.
	For $\lambda_V\so/\pi = 0.1, 0.3$, the critical chemical potential, compared to the \gls{gn} model result, increases when the temperature is decreased. 
	Increasing $\lambda_V$ further, leads to a change in this behavior, as mostly evident for $\lambda_V \so = \pi$ (visualized by the blue dot-dashed line).
	At this value of the vector coupling, the difference in the critical chemical potential compared to the GN model result, i.e., $\mu_c(\lambda_V, T) - \mu_c(0.0, T)$, is larger for, e.g., $T/\so = 0.4$ than for $T/\so = 0.0$.
	This results in a back-bending shape of the transition line, which reminds of a spinodal line in typical \gls{njl} or \gls{gn} model phase diagrams \cite{Buballa:2003qv}. 
	In this case, however, the left spinodal corresponds directly to the phase boundary line, since the phase transitions is of second order and the order parameter $\minHs(\mu, T)$ goes continuously to zero when crossing the boundary line from the \gls{hbp} to the \gls{sp}. 
	
\begin{figure*}[t]
		\centering
		\includegraphics{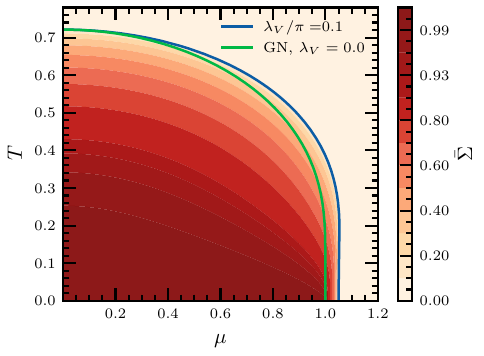}\hfill
		\includegraphics{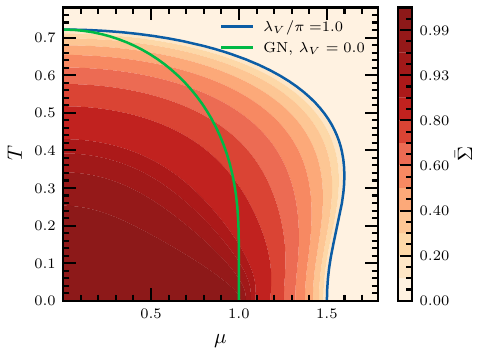}
		\caption{\label{fig:homCondPlot} Contour color maps in the $\left(\mu, T\right)$ plane for the value of the chiral condensate $\minHs(\mu, T)$ for \textbf{(left)} $\lambda_V \so / \pi= 0.1$, \textbf{(right)} $\lambda_V \so / \pi = 1.0$. The green lines represent the second order phase boundary of the $\left(2+1\right)$-dimensional \gls{gn} model, while the blue lines correspond to the phase boundary of the model \eqref{eq:FFmodel} for the respective value of $\lambda_{V}$. Continuous data for the contour plots is obtained using triangulation provided by \textit{Matplotlib} in \textit{Python3} \cite{Python3,4160265}. Note that the plot range in the $\mu$ and the $T$ axis differs from plot to plot in order to make the contour lines visible.}
	\end{figure*}

	The order of the phase transition becomes evident when studying the behavior of the order parameter $\minHs(\mu, T)$ as plotted in \cref{fig:homCondPlot}, where the value of $\minHw(\mu, T)$ for $\lambda_V \so / \pi = 0.1,1.0$ is visualized in a triangulated contour color map. 
	For all studied values of the vector coupling, one observes a continuous decrease of the chiral condensate when increasing the temperature starting within the \gls{hbp}.
	At $T=0$, we obtain $\minHs(\mu, T=0) = \so, \, \forall \mu \leq \so$ in consistency with the Silver Blaze property.
	When further increasing the chemical potential above $\so$, one again obtains a continuous decrease of the chiral condensate from $\minHs(\mu/\so = 1, T = 0) = \so$ to $\minHs(\mu = \mu_c, T = 0) = 0.0$ at the phase transition to the \gls{sp}. 
	This can be explained in context of the flatness of the effective potential of the \gls{gn} model at $\left(\mu/\so = 1, T = 0\right)$. 
	This flatness as a function of $\Hs$ is also present in the effective potential at $\lambda_V \neq 0$ at $\left(\mu/\so = 1, T = 0\right)$.
	However, the additional contribution due to the non-vanishing $\Hw$ causes solutions with higher densities to be favored, compare \cref{eq:hompot}.
	The coupling of the gap Eqs.~for $\minHs$ and $\minHw$, \cref{eq:GapSigma} and \cref{eq:GapOmega} respectively, leads then to a simultaneous decrease of $\minHs$ and a continuous increase of $\minHw$ when increasing $\mu$ for all chemical potentials $\mu \in [\so, \mu_c]$.  
	For all values of $\lambda_V$, the chemical potentials in this interval correspond to $\muSh = 1.0$, since this is the only point at zero temperature where the gap Eq.~\eqref{eq:GapSigma} allows for solutions $\minHs(\mu, T = 0.0)$ other than $\minHs / \so = 1.0$ or $\minHs / \so = 0.0$. 
	Due to the coupling of the gap Eqs.~for $\minHs$ and $\minHw$, the solution of the gap Eq.~for $\minHw(\mu, T = 0)$ can directly be read of the plot for this range of chemical potentials as it is given by $\minHw = \mu - \muSh_c(T=0) = \mu - \so$.
	The density is given by $\minHn = - \ii \minHw / \lambda_V$ due to the gap \cref{eq:GapOmega} or, equivalently, the Ward identity \eqref{eq:Ward} for $\omega_3$.
	
	In the \gls{gn} model, the density jumps from $0$ to $\so^2 / (2\pi)$ when crossing the phase transition towards the \gls{hbp} at zero temperature\footnote{Note that there is a factor of $2$ difference in the definition of the density between this work and \Rcite{Urlichs:2007zz}.} \cite{Urlichs:2007zz}. 
	This jump becomes a continuous transition for all $\lambda_V > 0.0$ and $\minHn / \so^2 = 1 / (2\pi)$ is reached in the models with vector interaction directly at $\mu = \mu_c$.
	All other values of $\minHn(\mu, T = 0) / \so^2 \in [0.0, 1 / (2\pi))$ are obtained when continuously increasing the chemical potential from $\mu / \so > 1.0$, whereas $\minHs(\mu, T=0)$ continuously decreases as discussed above.
	This continuous decrease for $\lambda_V > 0.0$ is enforced by the additional term $\propto - \Hn^2$ in the effective potential \eqref{eq:hompot}.
	The gap equation \cref{eq:GapOmega} allows a non-vanishing density only at $\mu / \so > 1.0$ and $\minHs / \so < 1.0$ such that at $\mu / \so = 1.0$ only the zero density solution is allowed.
	This is another clear indication that the phase boundary for $\lambda_V > 0.0$ is of second order, also at $T = 0$.

\begin{figure*}[t]
		\centering
		\includegraphics{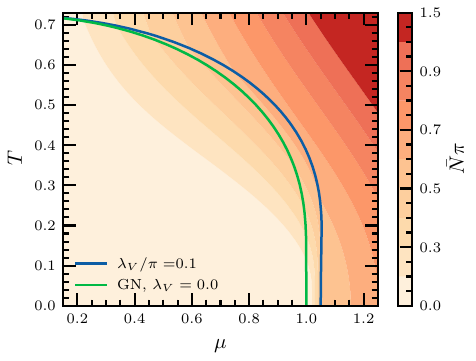}\hfill
		\includegraphics{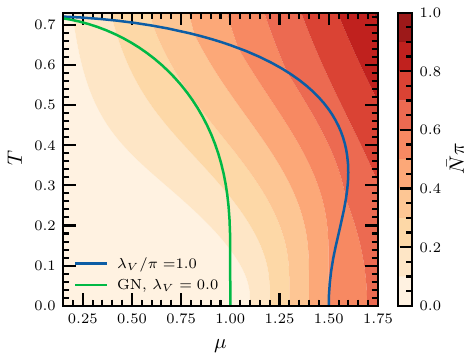}
		\caption{\label{fig:homDensPlot} Contour color maps in the $\left(\mu, T\right)$ plane for the value of the density $\minHn(\mu, T) \pi  = -\ii\minHw(\mu, T) \pi / \lambda_V $ for \textbf{(left)} $\lambda_V \so / \pi= 0.1$, \textbf{(right)} $\lambda_V \so / \pi = 1.0$.
			The green lines represent the second order phase boundary of the $\left(2+1\right)$-dimensional \gls{gn} model, while the blue lines correspond to the phase boundary of the model \eqref{eq:FFmodel} for the respective value of $\lambda_{V}$. Continuous data for the contour plots is obtained using triangulation provided by \textit{Matplotlib} in \textit{Python3} \cite{Python3,4160265}. Note that the plot range in the $\mu$ axis, the $T$ axis and the discrete color bar differs from plot to plot in order to make the contour lines visible. Also, the discrete contour levels are not necessarily linearily distributed.}
	\end{figure*}

	At non-vanishing temperatures the relation between the chemical potential and the value of $\minHw$ becomes non-trivial and has to be determined via the numerical solution of the gap Eqs.~\eqref{eq:GapSigma}, \eqref{eq:GapOmega} and finding the global minimum of the effective potential \eqref{eq:hompot}.
	\cref{fig:homDensPlot} depicts the density as a triangulated contour color map in the $\left(T, \mu\right)$ plane for $\lambda_V \so / \pi = 0.1, 0.5, 1.0$, respectively.
	Again, at $T=0.0$ we observe the Silver Blaze property as the density is zero for $\mu / \so  \in [0.0, 1.0]$.
	For non-vanishing temperature, this no longer holds and one observes the onset of the density for $\mu / \so < 1.0$. 
	In general, the density at fixed chemical potential grows monotonically with the temperature and vice versa.
	Comparing the phase diagrams for different vector couplings, one generally observes that the density is smaller for larger $\lambda_V$ when comparing at fixed values of $T$ and $\mu$.  
	
	Note that, in contrast to \gls{njl} model results \cite{Sasaki:2006ws}, one does never obtain a first-order phase transition in the $\left(\mu, T\right)$ plane.
	This is also relevant in the context of an \gls{ip}, that typically covers the region of a first order phase transition between the \gls{hbp} and \gls{sp} that is present when restricting to homogeneous phases.
	It is worth to note that there exists a first order phase transition at $\muSh / \so  = 1.0$ and $T / \so=0$ for any value of $\lambda_V$.
	This is caused by the effective potential of the \gls{gn} model being flat within $\Hs = [0.0, \so]$, but the additional contribution of $\minHw$ favors solutions of the gap Eqs.~with a high density leading to a jump of the global minimum from $\left(\minHs / \so, \minHw / \so \right) = \left(1.0, 0.0\right)$ at $(\muSh / \so=1.0-\epsilon, T / \so = 0.0)$ to  $\left(\minHs / \so, \minHw / \so\right) = \left(0.0, \ii \lambda_V \so / (2\pi)\right)$ at $(\muSh=1.0, T = 0.0)$, where $\epsilon$ can be infinitesimally small.
	This demonstrates that one has to be very careful about drawing $\left(\mu,T\right)$ phase diagrams, when using $\muSh$ as an external parameter in the computation and the chemical potential $\mu$ as a variable instead.
	As discussed in \cref{sec:methomcond}, the same value of $\left(\muSh, T\right)$ can correspond to multiple points in the $(\mu, T)$ plane or vice versa such that one has to always compare the corresponding values of the effective potential \eqref{eq:hompot}.
	
	\subsection{Stability analysis with mixing \label{sec:stab_sym}}
	
	\subsubsection{Symmetric phase}
	
	Within the \gls{sp} the Hessian $H_{\auxf_j \auxf_k}$ is diagonal for the original field basis, i.e., $\vdbf = \vauxf = \left(\sigma, \omega_\nu\right)$, since all non-vanishing off-diagonal elements are proportional to $\minHs$, see \cref{eq:Hessian} and the formulas in \cref{app:stab_analysis}.
	In this case, the analysis is similar to the one described in \cref{sec:stab_analysis}.
	The eigenvalues/diagonal elements of $H$ are the respective bosonic two-point vertex functions $\Gamma^{(2)}_{\dbf_{j}}$ of the fields $\left(\sigma, \omega_\nu\right)$.  
	Thus, they can be analyzed to study (the absence of) instabilities towards the \gls{ip} and the existence of the moat regime, similar to the analysis in \Rcite{Pannullo:2023one} without vector interactions. 
	
\begin{figure*}[t]
		\centering
		\includegraphics{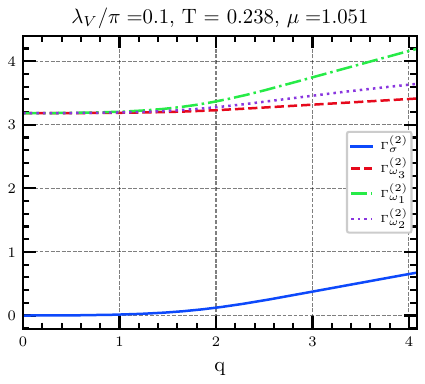}\hfill
		\includegraphics{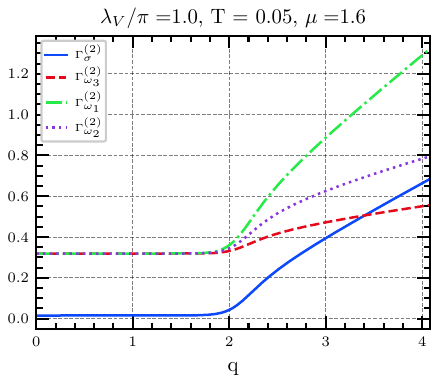}
		\caption{\label{fig:TwoPointSP} Bosonic two-point vertex functions $\Gamma^{(2)}_{\dbf_j}(q)$ as functions of the momentum of the perturbation $q$.
			\textbf{(left)} At the homogeneous second order phase transition for $\lambda_V \so / \pi = 0.1$. 
			\textbf{(right)} Within the \gls{sp} for $\lambda_V \so / \pi = 1.0$.
			Note the different plot ranges for the $y$ axis in the left and right panel.}
\end{figure*}

	In the left plot of \cref{fig:TwoPointSP}, the eigenvalues $\Gamma^{(2)}_{\dbf_{j}}$ of the Hessian are plotted as functions of the momentum $q$ of the perturbation $\ptdbf_j$ for $\lambda_V \so / \pi = 0.1$ and $(\mu/\so, T/\so) = \left(1.051, 0.238\right)$. 
	This point in the phase diagram lies directly on the second order phase boundary between the \gls{sp} and \gls{hbp} (compare \cref{fig:homboundary}). 
	One of the eigenvalues is zero at $q=0$ and corresponds to the order parameter $\sigma$ undergoing the phase transition.
	Since the two-point vertex functions $\Gamma^{(2)}_{\dbf_{j}}(q)$ can be interpreted as the curvature of the effective action \eqref{eq:effaction} in the direction of $\dbf_{j}(q)$, it is expected that this curvature goes to zero for $q=0$ and $\dbf_{j} = \sigma$ at the second-order homogeneous phase transition. 
	The field $\omega_\nu$ is not an order parameter and, thus, the corresponding bosonic two-point vertex functions do not show signals of the phase transition. 
	
	We always obtain that the two-point vertex functions $\Gamma^{(2)}_{\auxf_{j}}(q)$ are monotonically increasing functions as can directly be seen for $T=0$ by taking the zero temperature limit for all diagonal elements of the Hessian using the formulas in \cref{app:stab_analysis} and setting $\minHs = 0$.
	An example for the bosonic two-point vertex functions within the \gls{sp} is plotted on the right side of \cref{fig:TwoPointSP}.
	In the plot, we used the largest of the studied vector couplings $\lambda_V \so / \pi = 1.0$.
	A non-monotonic behavior of the two-point vertex function is never observed within in the \gls{sp}, which we studied for a large range of chemical potentials and temperatures.
	For $\Gamma^{(2)}_{\sigma}$, this result was already presented in \Rcite{Pannullo:2023one}, while in this work we also studied the two-point vertex functions $\Gamma^{(2)}_{\omega_\nu}(q)$, that correspond to the included vector interactions.
	In the \gls{sp}, however, the analysis yields the same conclusion as the one in \Rcite{Pannullo:2023one}.
	We do neither observe an instability towards an \gls{ip} nor a moat regime for all studied vector couplings.
	Together with the argument that all observed \glspl{ip} feature a second order phase boundary towards the \gls{sp} -- which would be detected by the stability analysis \cite{Koenigstein:2021llr} -- have so far never been observed, we consider the absence of such an instability a strong indication for the non-existence of an \gls{ip} within this model. 
	
	In the \gls{gn} model there still exists a degeneracy between inhomogeneous condensates and homogeneous phases at zero temperature as found with a particular ansatz function \cite{Urlichs:2007zz}. 
	This is also consistent with the stability analysis, where the bosonic two-point vertex function $\Gamma^{(2)}_\sigma(q)$ is flat and vanishes for a certain interval in $q$ is found at the point $\left(\mu /\so = 1.0, T / \so = 0\right)$ \cite{Pannullo:2023one} -- the same point in the phase diagram where also the homogeneous potential is flat.
	This flatness of the bosonic two-point vertex function also occurs at the critical chemical potential $\mu_c$ at any value of $\lambda_V$ at $T= 0.0$, again indicating a similar degeneracy between the \gls{sp} and the \gls{ip} as before.
	This property could already be guessed from the right plot in \cref{fig:TwoPointSP}, which is still at finite, but low enough temperature such that the two-point vertex function $\Gamma^{(2)}_\sigma$ is almost flat for small $q$. 
	We expect, however, that a degenerate condensate would not be given by the one-dimensional kink ansatz from \Rcite{Urlichs:2007zz} (see Figure 5.4 therein), since its density $\Hn$ remains homogeneous and is smaller than $\Hn / \so^2 = 1 / \left(2\pi\right)$, which is the density corresponding to $\minHs = 0.0$ when solving the gap Eqs.~\eqref{eq:GapSigma}, \eqref{eq:GapOmega} at $\left(\mu /\so = 1.0, T / \so = 0\right)$.
	Thus, the homogeneous solution $\left(\minHs / \so = 0.0, \minHw / \so  = \ii \lambda_V \so  / \left(2\pi \right) \right)$ is expected to be favored over the ansatz in \Rcite{Urlichs:2007zz}, as the bosonized action favors solution of the gap Eqs.~with higher densities, compare \cref{eq:bos_action} using $\minHw = \ii \lambda_V \Hn$. 
	
\subsubsection{Complex conjugate eigenvalues for $q=0$ \label{sec:stab_broken}}
Within the \gls{hbp}, one obtains mixing between $\sigma$ and $\omega_3$ when inspecting the static Hessian $H(q=0)$ as $H_{\sigma \omega_3}(q=0) \neq 0$.
When considering $q\neq 0$, there can also occur mixing involving the spatial components of $\omega_\nu$, see the discussion in \cref{sec:stab_analysis} and \cref{app:stab_analysis} for the Hessian matrix elements.
We start by focusing on the static Hessian $H(q=0)$.
To study these static mixing effects, the Hessian \eqref{eq:Hessian} can be considered only for $\auxf_j, \auxf_k \in \left\{\sigma, \omega_3\right\}$.
Perturbations about the spatial components of the vector field are not interesting in this case, since the static Hessian is diagonal anyway with respect to $\omega_1$ and $\omega_2$.
The main finding of this section is the observation of complex-conjugate eigenvalues of $H(0)$ in the certain regions within the \gls{hbp} through these mixing effects between $\sigma$ and $\omega_3$. As a consequences, we argue that the bosonic two-point correlation functions are oscillatory, but exponentially suppressed according to the analysis of the propagator poles presented in \cref{sec:compl_conj_Hessian}.  

\begin{figure}[t]
	\includegraphics{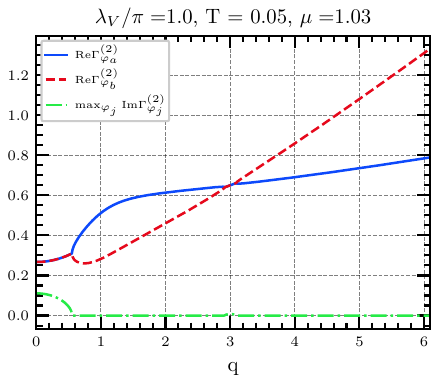}
	
	\caption{\label{fig:TwoPointCCStatic} The real and imaginary part of the bosonic two-point vertex functions $\Gamma^{(2)}_{\dbf_j}(q)$ are plotted as functions of the momentum of the perturbation $q$. The bosonic two-point vertex functions are obtained as eigenvalues of $H_{\auxf_j, \auxf_k}$ with $\auxf_j, \auxf_k \in \left\{\sigma, \omega_3\right\}$. Note that only $\Im\Gamma^{(2)}_{\dbf_a}(q)$ is plotted, since $\Im\Gamma^{(2)}_{\dbf_a} = -\Im\Gamma^{(2)}_{\dbf_b}$.} 
	\end{figure}

At any temperature and $\mu \neq 0$ within the \gls{hbp}, one obtains mixing between $\sigma$ and $\omega_3$ such that the physical basis $\dbf_j \neq \auxf_j$.
As extensively discussed in \cref{sec:compl_conj_Hessian}, this mixing can lead to complex-conjugate eigenvalue pairs of the static Hessian $H(0)$ depending on the parameters $\mu$ and $T$.
An example of this phenomenon is shown in \cref{fig:TwoPointCCStatic}, where the real and imaginary eigenvalues of the Hessian matrix $H_{\auxf_{j} \auxf_k}(q)$ with $\auxf_j, \auxf_k \in \left\{\sigma, \omega_3\right\}$ are plotted for $\left(\mu /\so = 1.03, T / \so = 0.05\right)$ and $\lambda_V\so / \pi = 1.0$.
The eigenvectors $\dbf_{a}$ and $\dbf_b$ are given by $q$-dependent linear combinations of $\sigma$ and $\omega_3$, i.e., $\dbf_j(q) = c_j(q) \sigma + d_j(q)\omega_3(q)$.
The non-vanishing imaginary part of the eigenvalues for $q=0$ decreases as a function of $q$ in this analysis such that real-valued eigenvalues are obtained for some relatively small $q / \so \approx 0.5$. 
Then, also the degeneracy $\Re \Gamma^{(2)}_{\dbf_a} = \Re \Gamma^{(2)}_{\dbf_b}$ is no longer enforced by $\mathcal C \mathcal K$ invariance (see the discussion in \cref{sec:compl_conj_Hessian}) resulting in an apparently non-analytic behavior of both two-point vertex functions at $q / \so \approx 0.5$ and $q / \so \approx 3.0$. 
It is interesting to note that whenever the real parts of $\Gamma^{(2)}_{\varphi_{a/b}}(q)$ are equal to each other one also observes a non-vanishing imaginary part of $\Gamma^{(2)}_{\varphi_{a}}(q) = -\Gamma^{(2)}_{\varphi_{b}}(q)$.
This is a result of the invariance under the $\mathcal C \mathcal K$ and the fact that both eigenvectors always need to fulfill $\varphi_a \neq \varphi_b$.
In the static case with $q=0$, the complex-conjugate eigenvalue pairs leads to bosonic propagators that are sinusoidal modulated alongside the usual exponential decay.
This follows as the inverse of the bosonic two-point correlation function is given by the two-point vertex function, which in the mean-field approach is given by the eigenvalues of the Hessian matrix when expanding about the thermodynamic (homogeneous) ground state, see \cref{sec:compl_conj_Hessian}.
A low-momentum expansion of the two-point vertex function then yields the described behavior through the appearance of roots of the propagator with a non-vanishing real and imaginary part, see the discussions in \cite{Nishimura:2014kla, Schindler:2019ugo, Schindler:2021cke} and \cref{sec:compl_conj_Hessian}.
Note that the behavior of the two-point vertex functions in \cref{fig:TwoPointCCStatic} can change at $q \neq 0$ when including perturbations about $\omega_1$ and $\omega_2$, as discussed above. 
Thus, \cref{fig:TwoPointCCStatic} should only be understood as an example for the effects of mixing and not as a full solution of the momentum dependence of the Hessian matrix $H(q)$ of the full model \eqref{eq:FFmodel}. 
It would be the full solution of the momentum dependence when only interactions proportional to $\sigma$ and $\omega_3$ would be studied.

\begin{figure*}[t]
\centering
\includegraphics{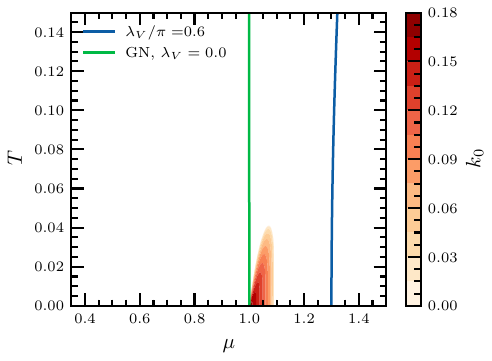}\hfil
\includegraphics{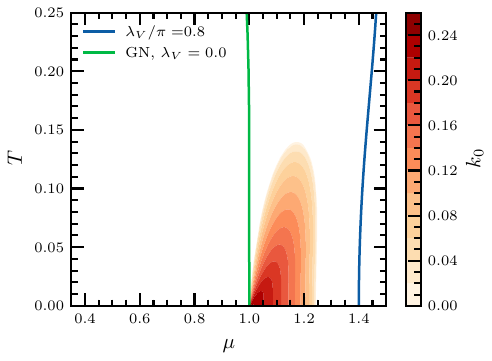}\hfil
\includegraphics{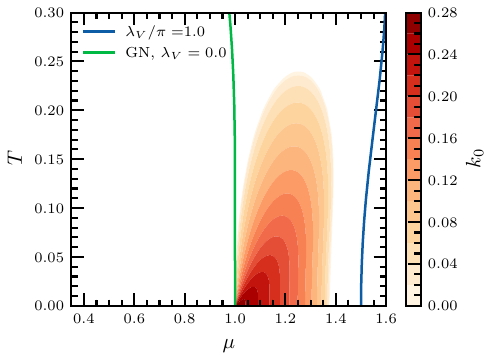}
\caption{\label{fig:koCC} Contour color maps in the $\left(\mu, T\right)$ plane for the value of the maximal imaginary part of the eigenvalues at $q=0$, denoted by $k_0$, for \textbf{(top left)} $\lambda_V \so / \pi= 0.6$, \textbf{(top right)} $\lambda_V \so / \pi = 0.8$, \textbf{(bottom)} $\lambda_V \so / \pi = 1.0$. The green lines represent the second order phase boundary of the $\left(2+1\right)$-dimensional \gls{gn} model, while the blue lines correspond to the phase boundary of the model \eqref{eq:FFmodel} for the respective value of $\lambda_{V}$. Continuous data for the contour plots is obtained using triangulation provided by \textit{Matplotlib} in \textit{Python3} \cite{Python3,4160265}. Note that the plot range in the $\mu$ axis, the $T$ axis and the discrete color bar differs from plot to plot in order to make the contour lines visible.  }
\end{figure*}

	To characterize the regime with spatial oscillatory behavior of propagators in the $(\mu, T)$ plane, we use the maximal imaginary part of the eigenvalues at $q= 0$ given by
	\begin{equation}
k_0 =  \max_{\dbf_j} \left(\Im \Gamma^{(2)}_{\dbf_j}(q=0)\right).
\end{equation}
This is an important scale for the momentum of the sinusoidal oscillation of the bosonic propagator, as a non-vanishing $k_0$ induces a non-vanishing real part of the propagator poles, see the discussion in \cref{sec:compl_conj_Hessian}.
In \cref{fig:koCC}, we plot $k_0$ in the $(\mu, T)$ plane using a color code for $\lambda_V \so / \pi \in \{0.6, 0.8, 1.0\}$.
A region with $k_0 \neq 0$ for chemical potentials $\mu / \so > 1.0 $ and rather small temperatures is observed for all three vector couplings.
We note that $\vdbf(q=0) = \left(z_1 \sigma + z_2 \omega_3, z_2^* \sigma + z_1^* \omega_3, \omega_1, \omega_2 \right)$ with complex-valued coefficients $z_1$ and $z_2$ for all studied points in the phase diagram and all studied vector couplings $\lambda_V$.
The extent of the region with complex-conjugate eigenvalue pairs both in $\mu$ and in $T$ direction strongly depends on the value of the vector coupling.
For $\lambda_V \so / \pi = 0.6$, this region's extent is significantly smaller than for $\lambda_V \so / \pi = 1.0$.
For all studied vector couplings, the width in the $\mu$ direction of the regime with oscillating propagator behavior decreases for larger temperatures, until $k_0$ goes to zero.
This is caused by thermal fluctuations suppressing the oscillatory behavior in the propagators, as can be seen from $k_0$ decreasing monotonically with the temperature at fixed $\mu$.
Such a behavior is typical for regimes with spatial oscillations, because the thermal fluctuations tend to destroy ordering in general \cite{Kolehmainen:1982jn,Lenz:2020bxk, Pisarski:2020dnx,Stoll:2021ori}.
For all three vector couplings, $k_0$ jumps from zero to a non-vanishing value when crossing $\mu/\so > 1.0$ at zero temperature in consistency with the Silver Blaze property.
This can also be derived from the formulas in \cref{app:stab_analysis} for the off-diagonal elements.
Investigations of $k_0$ as show that $k_0(\mu, T)$ seems to have a rapid but continuous onset from zero when increasing $\mu$ from the left of the regime with spatial oscillations at any $T \neq 0$. 
In the typical literature, the transition from $k_0 = 0$ to $k_0 \neq 0$ is also called disorder line \cite{Pisarski:2020dnx} -- as it is not a phase transition in the typical sense, but marks a distinctly different behavior of propagators. 

\begin{figure}[t]
\begin{center}
	\includegraphics{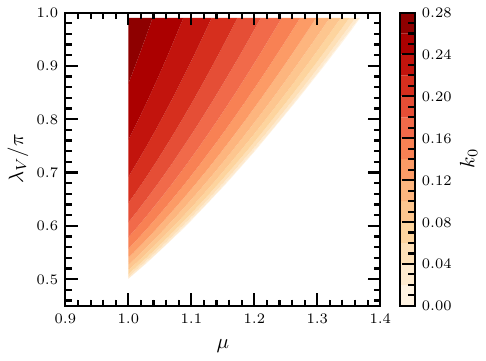}\hfil
	\caption{\label{fig:zeroTCCk0} Contour color maps in the $\left(\lambda_V, \mu\right)$ plane for the value of the maximal imaginary part of the eigenvalues at $q=0$, denoted by $k_0$, at zero temperature. Continuous data for the contour plots is obtained using triangulation provided by \textit{Matplotlib} in \textit{Python3} \cite{Python3,4160265}. }
\end{center}
\end{figure}

In \cref{fig:zeroTCCk0}, we plot $k_0$ at zero temperature as a function of $\lambda_V \so$ and $\mu/\so$ and observe the above described non-analytic behavior at all $\lambda_V \so > 0.5 \pi$. 
Precisely at $\lambda_{V,c} \so = 0.5\pi$, however, there is a continuous onset of $k_0$.
For $\lambda_{V} \so < 0.5 \pi$, we do not find $k_0 \neq 0$ at all.
Since the width of the region  with $k_0 \neq 0$ in the $\mu$ direction is largest at $T=0$ for all studied vector couplings, we expect that one finds $k_0 = 0$ for all vector couplings lower than $\lambda_{V,c}$ both at zero and non-zero temperature.
Thus, we expect there is no region with complex-conjugate eigenvalues appearing in the whole $\left(\mu, T\right)$ plane for $\lambda_V \so < 0.5 \pi$.
From \cref{fig:zeroTCCk0} it becomes clear that the extent of the region with complex-conjugate eigenvalues of $H(0)$ grows with increasing vector coupling when $\lambda_V >\lambda_{V,c}$.
Also, the value of $k_0$ grows monotonically when increasing $\lambda_V$ from any value larger than $\lambda_{V,c}$. 
Both observations show that the increase of the vector coupling increases the mixing effects, which is expected since an increase of $\lambda_V$ lowers the difference $H_{\sigma \sigma}(0) - H_{\omega_3 \omega_3}(0)$. 
This difference between the two diagonal elements needs to be smaller than the product of the off-diagonal elements such that the eigenvalues of the corresponding $2\times2$ block are complex-conjugate eigenvalue pairs. 
Thus, the increase of mixing effects also amounts to a growth of the extent of the regime with spatially oscillating propagators and the maximal obtained values of $k_0$.

Although $k_0$ is a useful observable to quantify the appearance of complex-conjugate eigenvalues, it is not the unique scale determining the oscillation of the propagator $G_{\dbf_{j}}$. 
Instead, the scales for the oscillation and the exponential decay of $G_{\dbf_{j}}$ are given by the respective real and imaginary parts of the roots $q^{1,2}_{\dbf_{j}}$ of \cref{eq:invprop_approx}. 
These roots are given by 
\begin{equation}
q^{1,2}_{\dbf_{j}} = \pm \ii \sqrt{\frac{\Gamma^{(2)}_{\dbf_{j}}(0)}{Z_{\dbf_{j}}}}. \label{eq:prop_poles}
\end{equation}
In order to compute these roots numerically, one can use the fact that $\dbf_{j}(q)$ has a weak dependence on $q$ around $q=0$. 
In practice, the computation of $Z_{\dbf_{j}}$ involves the discrete differentiation of the eigenvalues of $H$ (compare \cref{eq:invprop_approx} and \cref{eq:wavefunc_gen}).
Thus, its computation has to be performed by carefully taking into account the discretization error in $q$ and the change of basis $\dbf_{j}(q)$ in this area. 
Also, the imaginary part of $Z_{\dbf_{j}}$ is very small such that one also encounters problems with under-flowing of double precision.
In test runs of this evaluation the maximum value for $\Im Z_{\dbf_{j}}$ encountered was on the order of $10^{-3}$, while $\Re Z_{\dbf_{j}}$ is typically of order $10^{-1}$.
However, the problems involving the numerical evaluation of $Z$ make a precise determination of the poles using \cref{eq:prop_poles} impractical with respect to the study of the whole phase diagram.

Hence, we decided to use\footnote{Note that by definition of $\Gamma^{(2)}_\chi$ its imaginary part is always positive such that the determination of the argument in \cref{eq:r_def} is always valid.} 
\begin{equation}
\theta = \arg\left( \sqrt{\frac{\Gamma^{(2)}_\chi(0)}{ \Re Z_\chi}} \right) = \frac{1}{2} \arccos\left(\frac{\Re \Gamma^{(2)}_\chi(0) }{ |\Gamma^{(2)}_\chi(0)|}\right) \label{eq:r_def}
\end{equation}
with \begin{equation}
\chi = \argmax_{\dbf_j} \left(\Im \Gamma^{(2)}_{\dbf_j}(q=0)\right)
\end{equation}
in order to compare the scale of oscillation in comparison to the exponential decay.
\cref{eq:r_def} gives the complex argument $\theta$ of $q^{1,2}$ when setting the imaginary part of $Z_{\dbf_{j}}$ to zero. 
As argued above, the error of this approximation should be rather small.
Note that the numerical computation of $\theta$ using \cref{eq:r_def} does require an evaluation of $Z_{\dbf_{j}}$. 

\begin{figure*}[t]
\begin{center}
	\includegraphics{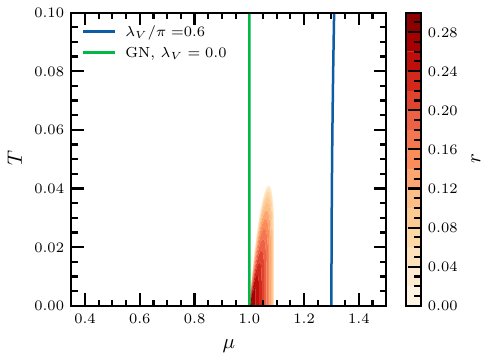}
	\includegraphics{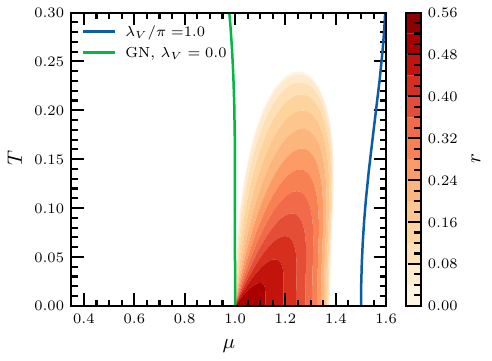}
\end{center}
\caption{\label{fig:ComplexArg} Contour color maps in the $\left(\mu, T\right)$ plane for the ratio $r$ between the frequency of the spatial oscillation and the exponential decay rate of the exponential decay of the propagator. \textbf{(left)} $\lambda_V \so / \pi= 0.6$,  \textbf{(right)} $\lambda_V \so / \pi = 1.0$. The green lines represent the second order phase boundary of the $\left(2+1\right)$-dimensional \gls{gn} model, while the blue lines correspond to the phase boundary of the model \eqref{eq:FFmodel} for the respective value of $\lambda_{V}$. Continuous data for the contour plots is obtained using triangulation provided by \textit{Matplotlib} in \textit{Python3} \cite{Python3,4160265}. Note that the plot range in the $\mu$ axis, the $T$ axis and the discrete color bar differs from plot to plot in order to make the contour lines visible.}
\end{figure*}

We find $r = \tan \theta = \Re q^{1,2}_\chi / \Im q^{1,2}_\chi  $ as the ratio of the oscillation frequency $\Re q^{1,2}_\chi$ and the decay rate $\Im q^{1,2}_\chi$.
In \cref{fig:ComplexArg}, we plot $r$ for two different values $\lambda_V \so = 0.6 \pi$ and $\lambda_V \so = \pi$.
The maximal value of $r$ obtained is $0.58$ for $\lambda_V \so = \pi$ such that frequency of the oscillation is larger than half of the exponential decay rate at this point in the phase diagram. 
Similar to the behavior of $k_0$, we find that for smaller vector couplings lower values of $r$ are obtained in general.   
This is expected, since mixing effects will not be as drastic for lower vector couplings, see the discussion above.
For $\lambda_V \so = 0.6\pi$, the maximal value of $r$ is roughly $0.3$.
In general, the obtained contour lines for $r$ are very similar to the ones of $k_0$, see \cref{fig:koCC}.
The figure demonstrates that for large parts of the regime with oscillatory behavior the wavelengths are in the order of the inverse of the exponential decay rate.

\subsubsection{Complex-conjugate eigenvalue pairs emerging at $q\neq0$}

\begin{figure}[t]
\centering
\includegraphics{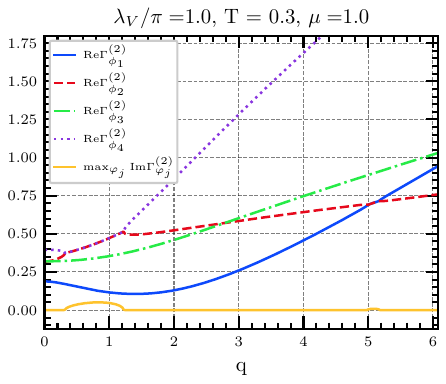}
\caption{\label{fig:TwoPointCCDyn} The real and imaginary part of the bosonic two-point vertex functions $\Gamma^{(2)}_{\dbf_j}(q)$ as functions of the momentum of the perturbation $q$. The bosonic two-point vertex functions are obtained as eigenvalues of $H_{\auxf_j, \auxf_k}$ with $\auxf_j, \auxf_k \in \left\{\sigma, \omega_\nu\right\}$. Note that the non-vanishing imaginary part does not necessarily belong to similar eigenvectors $\dbf_{j}(q)$ and complex-valued two-point vertex functions appear as complex-conjugate pairs.} 
\end{figure}

Already in \cref{fig:TwoPointCCStatic}, where only mixing of $\sigma$ and $\omega_3$ was taken into account, one can observe the appearance of complex-conjugate eigenvalues of the Hessian at $q\neq 0$.
However, considering a $4\times4$ Hessian matrix $H_{\auxf_{j} \auxf_k}(q)$ with $\auxf_j, \auxf_k \in \left\{\sigma, \omega_\nu\right\}$ yields a more involved mixing pattern, since also $\omega_1$ contributes to mixing effects with $\sigma$ when studying $q \neq 0$.  
The other vector component $\omega_2$ is not mixing with $\sigma$ (but with the other components of $\omega_\nu$), since we choose $\mathbf{q} = \left(q, 0\right)$ in \cref{sec:stab_analysis}.
The roles of $\omega_1$ and $\omega_2$ are exchanged if we choose $\mathbf{q}$ to be aligned with the $x_2$ axis. 
Since the stability analysis turns out to be invariant under spatial rotations (see the discussion in \cref{sec:stab_analysis}), the eigenvalues are independent of the chosen spatial direction $\mathbf{q} / |\mathbf{q}|$. 

The more complicated mixing pattern is depicted in \cref{fig:TwoPointCCDyn}. 
From the plot one obtains that $H(0)$ has real eigenvalues, but then develops complex-conjugate eigenvalue pairs at some value of $q = q_B$, where we define 
\begin{equation}
q_B= \min_{q\in C}{q}, \quad C = \{q \in [0, \infty) | \mathrm{Im} \Gamma^{(2)}_{\phi_j}(q) \neq 0\}. \label{eq:qB_Def}
\end{equation}
In this case, one obtains that only two of the four eigenvalues have non-vanishing imaginary parts for fixed $q$, while the other two are real-valued.
As can also be seen from \cref{fig:TwoPointCCDyn}, complex-conjugate eigenvalue pairs can be obtained for multiple intervals in $q$.
This leads to the rather complicated behavior of two-point vertex functions with the real-parts of different eigenvalues becoming degenerate depending on the value of $q$.
Since complex-conjugate eigenvalue pairs can occur for all of the eigenvalues $\Gamma^{(2)}_{\dbf_{j}}$ with $j \in \{1, 2, 3, 4\}$, the yellow line in the plot always only describes the appearance of the maximal imaginary part in any of those eigenvalues.
Since the eigenvectors of $H$ can be strongly $q$-dependent especially for large $q$, one might even argue that the association of the eigenvalues using functions $\Gamma^{(2)}_{\dbf_{j}}(q)$ is not very insightful.
Nevertheless, \cref{fig:TwoPointCCDyn} certainly demonstrates the involved mixing effects between scalar and vector modes. 
Also it shows that there is certainly non-monotonic behavior of the real parts of the two-point vertex functions such that one cannot exclude the appearance of moat regime.
However, also the bosonic wave-function renormalization $Z_{\dbf_{j}}$ can become complex, see \cref{eq:wavefunc_gen}, and might not be a decent criterion for a moat regime where $\argmin_q \Re \Gamma^{(2)}_{\dbf_{j}}(q) \neq 0$.
This is caused by non-analytic behavior of $\Gamma^{(2)}_{\dbf_{j}}(q)$  around the regions with non-vanishing imaginary parts leading also to non-monotonic behavior.
For these reasons we refrain from studying the moat regime within the \gls{hbp} and focus on the complex-valued eigenvalues instead.  

\begin{figure*}[t]
\centering
\includegraphics{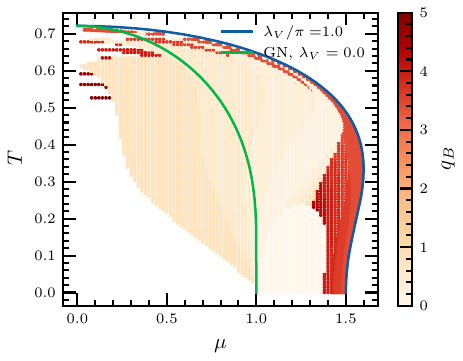}\includegraphics{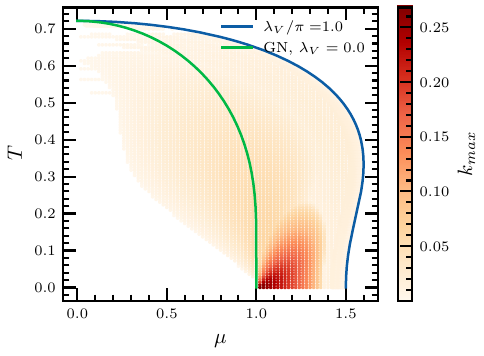}
\caption{\label{fig:4x4koCC} Contour color maps in the $\left(\mu, T\right)$ plane for $\lambda_V \so / \pi = 1.0$ encoding \textbf{(left)} $q_B$ (see \cref{eq:qB_Def}) \textbf{(right)} $\kmax$ (see \cref{eq:kmax_Def}). The green lines represent the second order phase boundary of the $\left(2+1\right)$-dimensional \gls{gn} model, while the blue lines correspond to the phase boundary of the model \eqref{eq:FFmodel}. Note that the plot range in the discrete color bar differs from plot to plot. The color bar for $q_B$ is cut off at $q_B / \so = 4.0$ such that the behavior of lower $q_B$ could be visualized accurately. The discretization of $q$ is given by $\Delta q = 0.2$ resulting in discretization errors in the computation of $q_B$ and $\kmax$.}
\end{figure*}

In the left plot of \cref{fig:4x4koCC}, we plot a color map in the $(T,\mu)$ plane encoding $q_B$ for $\lambda_V \so / \pi$. 
In the right plot of this figure we visualize    
\begin{equation}
\kmax = \max_{j, q} \left(\mathrm{Im} \Gamma^{(2)}_{\phi_j}(q)\right) \label{eq:kmax_Def}
\end{equation}
also using a color code.
\cref{fig:4x4koCC} demonstrates that complex-conjugate eigenvalues appear in large parts of the \gls{hbp} except for a rather small region at small temperatures and chemical potentials. 
For some parts of this region the obtained imaginary parts are rather small.
In consistency with the Silver Blaze property, $\kmax = 0$ for $T = 0$ and $\mu / \so < 1.0$.
For rather small chemical potentials, $k_\mathrm{max} / \so$ is of the order of $10^{-3}$ and $q_B / \so \gg 1$. 
Accordingly, the interpretation in terms of oscillating propagators as in the static case is not possible as a low-momentum expansion is not meaningful.
However, if existent, any oscillating effects in this region should be negligible anyhow given that any relevant imaginary part should be smaller than $k_\mathrm{max} / \so \sim 10^{-3}$, 
For larger chemical potentials, one obtains $q_B / \so < 1.0$ making a low-momentum expansion of the inverse propagator more sensible.
We note that due to the computational demands of computing multiple momentum integrals for the determination of the matrix entries of $H(q)$ and its diagonalization, the resolution in $q$ for the computation of the data \cref{fig:4x4koCC} was chosen as $\Delta q = 0.2$.
The rather coarse resolution in $q$ results in inaccuracies in the determination of $q_B$ and $\kmax$, because the intervals, where complex-valued eigenvalues occur, can be smaller than $\Delta q$.
This is evident for some data points around the homogeneous phase boundary as well as small $\mu$ and $T / \so \in [0.5, 0.7]$, where $q_B$ appears to have jumps when changing $\mu$ or $T$.
At these data points, it is likely that for some intervals in $q$, which are smaller than $\Delta q$, complex-valued $\Gamma^{(2)}_{\dbf_{j}}(q \neq 0)$ appear, that where missed such that the correct $q_B$ differs from the depicted data point.
Anyhow, the value of $\kmax$ is likely to be small anyhow withing these intervals.
Still, it is unclear whether complex-conjugate eigenvalues appearing at $q \neq 0$ have any consequence on the behavior of propagators. 

Overall, to the best of our knowledge there is no clear physical explanation for the appearance of complex-conjugate two-point vertex function pairs appearing at $q \neq 0 $ in the literature. 
Also, this work marks the first observation of this phenomenon in the literature -- again to the knowledge of the author.
For small $q_B$, a similar interpretation as in the static case of correlation functions with sinusoidal modulations might be meaningful, again performing a low-momentum expansion of the inverse propagators $G_{\dbf_{j}}^{-1}$, compare \cref{eq:invprop_approx} around some non-vanishing $q$ and determining propagator poles similar to \cref{eq:prop_poles}.
However, in the regions with larger $q_B / \so > 1.0$ such an expansion is certainly not sensible. 
An inversion and four-transformation of the obtained $\Gamma^{(2)}_{\dbf_{j}}$ to compute $G_{\dbf_{j}}(x,y)$ within these regions yields an ordinary exponential decay in the \gls{hbp}.
This is expected, since $\kmax$ is rather small compared to the real-part of the two-point vertex functions in these parameter regions (see \cref{fig:4x4koCC}) such that the effect of this phenomenon is negligible when studying bosonic two-point correlation functions.

\section{Implications for general four-fermion models in $\left(2+1\right)$ dimensions\label{sec:impl}}
In this section, we will argue that the previous analysis and, consequently, the phase diagram of the simpler model \eqref{eq:FFmodel} is identical to $\left(2+1\right)$-dimensional \gls{ff} models with all kind of local \gls{ff} interactions.
The general prototype of these models is defined by the action 
\begin{widetext}
\begin{align}
	&\mathcal{S}_{\mathrm{FF}}[\bar{\psi},\psi]   \label{eq:FFmodel_gen}   
	= \int_0^\beta \! \dr \tau \int\! \dr^2 \xstV   \left\{ \bar{\psi}\left(\slashed{\partial}+ \gamma_3 \mu \right)  \psi -  \sum_{j=1}^{16}  \left[  \tfrac{\coupling_S}{2 \N } \left(\bar{\psi}\,  \cm_j\, \psi\right)^2 + \tfrac{\coupling_V}{2 \N } \left( \left(\bar{\psi}\, \ii \cm_j  \gamma_3 \, \psi\right)^2 + \left(\bar{\psi}\, \ii \cm_j \vec \gamma \, \psi\right)^2 \right)  \right]\right\}, 
\end{align}
\end{widetext} 
where $\psi$ now contains $2\N$ four-component spinors (due to an additional "isospin" degree of freedom compared to the fields in \cref{eq:FFmodel}). The interaction vertices $c_{j,\nu}$ are $8\times8$ matrices in isospin and spin space and elements of  
\begin{equation}
\intchannelSet = \left(\cm_j\right)_{j=1,\ldots, 16} \equiv  \left(1, \ii\gamma_4, \ii\gamma_5, \gamma_{45}, \vpauli, \ii\vpauli\gamma_4, \ii\vec{\tau}\gamma_5, \vpauli\gamma_{45}\right), \label{eq:dirac_basis}
\end{equation}
where $\vpauli$ is the vector of Pauli-matrices acting on the isospin degrees of freedom. The matrices $\gamma_{4}$ and $\gamma_5$ anti-commute with the $\gamma_\nu$, while $\gamma_{45} \equiv \ii \gamma_{4} \gamma_5$, consequently, commutes with the $\gamma_\nu$. 
This model is invariant under a global $\U(4\N)$ chiral symmetry (see appendix A of \Rcite{Pannullo:2023one} for details). 

When neglecting vector interactions by setting $\lambda_V = 0$, this model has already been studied with respect to \glspl{ip} and the moat regime in \Rcite{Pannullo:2023one}. 
Also therein, it was shown that the \gls{ff} model \eqref{eq:FFmodel_gen} can easily be generalized to a Yukawa model by including local self-interactions and a kinetic term for the bosonic fields (after bosonization similar to \cref{eq:bos_action}) without altering the above analysis, as was also done in \Rcite{Pannullo:2023one}.
We refrain from doing this generalization to a Yukawa model for the action \eqref{eq:FFmodel_gen}.
This would further complicate the analysis through the medium induced mixing of vector and scalar modes, which would also affect the purely bosonic terms.  

For the \gls{ff} model \eqref{eq:FFmodel_gen} and any \gls{ff} model with a subset of its interaction channels, we will now demonstrate that their Hessian is block-diagonal with $4\times 4$ blocks that are identical to the $4\times4$ Hessian matrix \eqref{eq:Hessian} of the simpler model \cref{eq:FFmodel}.
Therefore, \cref{eq:FFmodel_gen} is bosonized, as before, with auxiliary bosonic fields $\left(\auxf_a, v_{a,\nu}\right)$. 
These fulfill the Ward identities 
\begin{equation}
\langle \auxf_a \rangle = - \frac{\lambda_S}{\N}\langle \bar{\psi} c_a \psi \rangle, \quad \langle v_{a,\nu} \rangle = - \ii \frac{\lambda_V}{\N}\langle \bar{\psi} c_a \gamma_\nu \psi \rangle.
\end{equation}
Similar to the discussion in \cref{sec:model}, one can assume that $\bar{v}_{a, j} = 0$ for $j=1,2$, leaving invariance under spatial rotation intact, where $\bar{v}_{a, j}$ are the homogeneous condensates of the vector fields.
Then, the derivation of the Hessian is similar to the one described in \Rcite{Pannullo:2023one} and in \cref{sec:stab_analysis}.
Through the $\U(4\N)$ global symmetry transformations, one can choose the values of $\auxf_a = \Hs \delta_{a,0}$ such that there is no mixing between those fields, i.e., $H_{\auxf_a \auxf_b} = 0$ for $a \neq b$. 
A similar transformation can be made for the fields $v_{a, \nu}$, such that $H_{v_{a,\nu} v_{b, \nu}} = 0$ for $a \neq b$.
With inspection of the trace in \cref{eq:Hessian}, one can then already infer that $H_{\auxf_{a,\nu} v_{b, \nu}} = 0$ for $a \neq b$.
Thus, one obtains mixing only for each respective $4\times 4$ block in $H$ with $a=b$, i.e., only $H_{\auxf_{a,\nu} v_{a, \nu}}$, $H_{\auxf_{a,\nu} \auxf_{a, \nu}}$ and $H_{v_{a,\nu} v_{a, \nu}}$ can be non-vanishing.
Again, inspecting the trace in \cref{eq:Hessian} and using the exchange properties of $c_j \in C$ with each other and the $\gamma_\nu$, one can derive that the matrix elements of these respective $4\times 4$ blocks in $H$ are identical to the $4\times 4$ Hessian matrix of \cref{eq:FFmodel} given by \cref{eq:Hessian}, respectively.
Thus, the Hessian of \cref{eq:FFmodel_gen} is block-diagonal consisting of $16$ matrices of size $4\times 4$, which are identical to the Hessian of the model studied in the previous sections.
Consequently, also the phase diagram is identical to the one of \cref{eq:FFmodel} with the enhancement of the \gls{hbp} when increasing $\lambda_V$, the absence of an \gls{ip} and the existence of regimes with complex-conjugate eigenvalues of $H$.

With this analysis and the results in \Rcite{Pannullo:2023cat}, we have demonstrated the absence of an \gls{ip} in $\left(2+1\right)$-dimensional \gls{ff} models with all kind of local interaction terms. 
Instead, we found a regime with spatially oscillating, but exponentially damped bosonic correlation functions when mixing between scalar and vector modes is allowed and the vector coupling exceeds a certain value of $\lambda_{V,c} \so = 0.5 \pi$, see, e.g., \cref{fig:zeroTCCk0}  and \cref{fig:ComplexArg} as well as the discussion in \cref{sec:stab_broken}.

\section{Conclusions \label{sec:concl}}
In this work, we analyzed the stability of homogeneous ground states for a $\left(2+1\right)$-dimensional \gls{ff} model \eqref{eq:FFmodel} with both scalar and vector interactions with particular emphasis on the effects of mixing between scalar and vector modes. 
The analysis was performed in the mean-field approximation, i.e., neglecting bosonic quantum fluctuations.
Also, we showed that our findings hold for a more general \gls{ff} model \eqref{eq:FFmodel_gen} consisting of all relevant interaction channels in $\left(2+1\right)$ dimensions.
We have shown the stability of homogeneous ground states against inhomogeneous perturbations and argued that this is strong evidence for the absence of an \gls{ip} in all $\left(2+1\right)$-dimensional fermionic theories with local \gls{ff} interactions.
Instead, a regime with spatially oscillating, but exponentially damped mesonic correlation functions has been detected through the appearance of complex-conjugate eigenvalues in the Hessian matrix for static perturbations $H(q=0)$. 
This regime is often also termed `quantum pion liquid' or `quantum spin liquid' \cite{Pisarski:2020dnx} -- in analogy to condensed matter literature -- and is directly related to the invariance of our model under $\mathcal{C} \mathcal{K}$ transformation at $\mu \neq 0$ \cite{Schindler:2019ugo, Schindler:2021cke}.
We also find regions with complex-conjugate eigenvalues in the Hessian matrix for perturbations with a non-vanishing momentum, i.e., $H(q =q_B \neq 0)$, while $H(0)$ has real-valued eigenvalues. 
So far, the implications of this phenomenon are not discussed in the literature.
However, we argued that the value of the obtained imaginary parts are so small that the mesonic correlation functions will follow an ordinary exponential decay within these regions.
The whole phase diagram of the \gls{ff} model \eqref{eq:FFmodel} was presented for different values of the vector coupling including the regimes with `quantum pion liquid' behavior -- which are located at low $T$ and intermediate $\mu$ within the \gls{hbp}. 
The typical ratio of the frequency of the oscillation and the exponential decay rate of the mesonic propagators is such that the oscillatory behavior could have an effect on related observables. 

The finding of the `quantum pion liquid' is to the best of our knowledge the first finding of this regime generated by the interplay of a attractive, scalar and a repulsive, vector \gls{ff} interaction -- apart from the report of this phenomenon in \Rcite{Haensch:2023sig}, where it was not the main focus of the study.
Since the `quantum pion liquid' is closely related to the invariance under the $\mathcal{C} \mathcal{K}$ operation \cite{Nishimura:2014kla}, we think this regime could also be of relevance in the phase diagram of \gls{qcd} at non-vanishing densities.
However, this work should be understood as a first, qualitative investigation of the underlying mechanism through mixing of scalar and vector modes.
The used $\left(2+1\right)$-dimensional models are too simplistic to make quantitative predictions for the phase diagram of \gls{qcd}.
We also want to note that the authors of \Rcite{Nishimura:2014kla} also report non-vanishing complex-conjugate eigenvalue pairs appearing in their static Hessian matrix (compare Figs.~17-21) in a P\gls{njl} model. 
However, these non-vanishing imaginary parts are generated by an analysis of Polyakov-loop effects and the mixing contributions stemming from fermionic point-like interactions are not included.
In \gls{qcd}, the contributions of both the gluon effects and of the mixing between the chiral condensate and vector mesons will play an important role and might lead to the existence of a `quantum pion liquid' regime at non-vanishing densities. 

There are several possibilities to extend on the present work.
A straightforward extension would be to study a $\left(3+1\right)$-dimensional \gls{qm} model including Polyakov-loop effects as well as vector meson interactions as in \Rcite{Haensch:2023sig}.
This allows to study the mixing of Polyakov-loop, chiral condensate and vector mesons in the same model.
Also, one could make phenomenologically more relevant predictions with respect to the parameter regions, where the `quantum pion liquid' should appear.
Moreover, it would certainly be interesting to study mixing effects beyond the mean-field approximation.
Thereby, it is interesting to note that already in \Rcite{Hands:2003dh, Strouthos:2003js} oscillating mesonic correlation functions have been measured in a $\left(2+1\right)$-dimensional \gls{gn} model, i.e., without including vector interactions, at finite number of fermion flavors $\N$ using lattice field theory.
In \Rcite{Pannullo:2023one}, we argued that these findings are reminiscent from an \gls{ip} that exists within the mean-field approximation at finite lattice spacing \cite{Buballa:2020nsi, Narayanan:2020uqt}.
However, as discussed in \Rcite{Cohen:1983nr,Lenz:2020bxk}, the naive and, following a similar argumentation, also the staggered fermion discretization yields off-diagonal interaction terms generated by the doublers, that contain $\gamma$ matrices. 
Thus, it is up to speculation whether the oscillating mesonic correlation functions observed in \Rcite{Hands:2003dh, Strouthos:2003js} might as well be generated by mixing effects as discussed in the present work.
When going beyond the mean-field approximation using a model that directly contains diagonal vector interactions as \cref{eq:FFmodel}, complex weights appear in the path integral \cite{Alexandru:2016ejd, Mori:2017zyl}.
This likely complicates using $1/N$ expansion techniques \cite{Klimenko:1993iz, Kneur:2007vj} or other analytical approaches like non-abelian bosonization \cite{Ciccone:2022zkg, Ciccone:2023pdk}.
Thus, one would have to either rely on Lefshetz thimble approaches \cite{Cristoforetti:2012su, Mori:2017zyl} or Functional Renormalization Group techniques \cite{Fu:2019hdw}.

Certainly, regimes where mesonic correlation functions are oscillatory will have effects on, e.g, the propagation of pions in heavy-ion-collision experiments.
Regarding experimental observables, direct consequences of spatially oscillatory regimes have to be worked out. 
Using the moat regime dispersion relation, this has been recently done using Hambury, Brown and Twist interferometry \cite{Rennecke:2023xhc, Fukushima:2023tpv} leading to signal peaks in two-particle correlation functions.
To study consequences for experimental observables in the `quantum pion liquid' regime one would first have to work out the phenomenologically relevant scales of the oscillation and exponential decay, e.g., using a Polyakov-loop \gls{qm} model as described above.

\section{Acknowledgments}
I thank M.~Buballa, T.~Motta, Z.~Nussinov, L.~Pannullo, R.~Pisarski, F.~Rennecke, D.~H.~Rischke, S.~Schindler, M.~Ogilvie, M.~Wagner for fruitful discussions related to this work.
Especially, I thank F.~Rennecke, S.~Schindler and M.~Ogilvie for help with the interpretation of complex-conjugate eigenvalues appearing in the Hessian matrix.
Furthermore, I like to thank A.~Koenigstein, L.~Pannullo, R.~Pisarski, S.~Schindler and M.~Wagner for helpful comments on this manuscript.
I acknowledge the support of the \textit{Deutsche Forschungsgemeinschaft} (DFG, German Research Foundation) through the collaborative research center trans-regio  CRC-TR 211 ``Strong-interaction matter under extreme conditions''-- project number 315477589 -- TRR 211.	
I acknowledge the support of the \textit{Helmholtz Graduate School for Hadron and Ion Research}.	

All numerical results in this work were obtained using $\textit{Python3}$ \cite{Python3} with various libraries \cite{Virtanen:2019joe, Harris_2020}.
The plots were designed using the \textit{Matplotlib} package \cite{4160265}.

\appendix
\section{Conventions for the Wick rotation \label{app:wick}}
In $\left(2+1\right)$-dimensional spacetime with the metric $\eta_{\mu \nu} = \mathrm{diag}(1, -1, -1)$ we demonstrate the Wick rotation using the action and partition function
\begin{align}
\S_{M, \mathrm{mix}}[\bar{\psi},\psi]   \label{eq:FFmodel_Mink}   
&= \int \dr x_0 \int \dr^2 \xstV   \Big\{ \bar{\psi}\left(\ii \gamma_M^\nu \, \partial_\nu \right)  \psi + \left.\left[  \tfrac{\coupling_S}{2 \N } \left(\bar{\psi}\,\psi\right)^2 + \tfrac{\coupling_V}{2 \N } \left( \bar{\psi}\, \ii  \gamma^\nu_M \, \psi\right) \left( \bar{\psi}\, \ii  \gamma_{M, \nu} \, \psi\right)    \right]\right\},  \\  Z &= \int \mathcal{D} \bar{\psi} \mathcal{D} \psi\, \e^{\ii \S_{M,\mathrm{mix}}[\bar{\psi},\psi]}
\end{align}
with most quantities as defined below \cref{eq:FFmodel}. The Gamma matrices $\gamma^\nu_M$ fulfill $\{\gamma^\mu_M, \gamma^\nu_M\} = 2 \eta_{\mu \nu}$, as usual. 
For the Wick rotation we use the conventions
\begin{align}
\tau = \ii x_0, \quad, \gamma_3 = \gamma^0_M ,\quad  \gamma_i = \ii \gamma_M^i, \, i \in  \{1,2\},   
\end{align}
such that one obtains $\left(\gamma_\nu\right)^2 =  1$.
This changes the vector \gls{ff} interaction term in the following way
\begin{align}
\left( \bar{\psi}\, \ii  \gamma^\nu_M \, \psi\right) \left( \bar{\psi}\, \ii  \gamma_{M, \nu} \, \psi\right) &= \left( \bar{\psi}\, \ii  \gamma^0_M \, \psi\right)^2 - \left( \bar{\psi}\, \ii  \gamma^i_M \, \psi\right)^2 = \left( \bar{\psi}\, \ii  \gamma_3 \, \psi\right)^2 + \bar{\psi}\left(\,\ii \gamma_i \, \psi\right)^2   
\end{align}
where the additional factor $(-\ii)^2 = -1$. With the other standard changes of the action and the definition $\S_{M,\mathrm{mix}} = \ii \S_{M, \mathrm{mix}}$ one obtains the Euclidean action \cref{eq:FFmodel}.

In these conventions the density $n$ is given by 
\begin{equation}
n = \frac{1}{\N} \frac{\dr \ln Z}{\dr \mu} =  -\frac{\langle \bar{\psi} \gamma_3 \psi\rangle}{\N}, 
\end{equation}
such that the Ward identity for $\omega_3$ \eqref{eq:Ward} follows. Note that these conventions and the Ward identity may differ from recent works, such as \Rcite{Mori:2017zyl}. 

\section{Formulas for the stability analysis with mixing \label{app:stab_analysis}}
In this section, we collect formulas needed for computation of the Hessian matrix elements $H_{\auxf_j, \auxf_k} = H_{\auxf_k, \auxf_j} $ with $\vauxf = \left(\sigma, \omega_\nu\right)$, compare \cref{eq:Hessian}.
Inserting 
\begin{equation}
S(\nu_n, \mathbf{p}) = \frac{-\ii \slashed{\tilde{p}} + \minHs}{\tilde{p}^2 + \minHs^2}, \, \tilde{p} = \left( \nu_n - \ii \muSh, \mathbf{p}\right)^T
\end{equation}
and the respective vertex $\vec{c} = \left(1, -\gamma_3, \ii \gamma_1, \ii \gamma_2\right)$ for $\vauxf = \left(\sigma,  \omega_3, \omega_1, \omega_2\right)$ in \cref{eq:Hessian}, respectively, one obtains for the diagonal elements
\begin{widetext}
\begin{align}
	H_{\sigma \sigma} &= \frac{1}{\lambda_S} + \frac{N_\gamma}{\beta} \int \frac{\dr^2 p}{(2\pi)^2} \sum_{n=-\infty}^{\infty} \frac{- \tilde{p}^2 - \mathbf{p} \mathbf{q} + \minHs^2}{(\nu_n - \ii \muSh)^2 + (\mathbf{p} + \mathbf{q})^2 + \minHs^2} \frac{1}{(\nu_n - \ii \muSh)^2 + \mathbf{p}^2 + \minHs^2}, \label{eq:H_ss} \\
	H_{\omega_\nu \omega_\nu} &= \frac{1}{\lambda_S} + (1 - 2\delta_{\nu,3}  ) \frac{N_\gamma}{\beta} \int \frac{\dr^2 p}{(2\pi)^2} \sum_{n=-\infty}^{\infty} \frac{(2\delta_{\nu,\alpha} - 1  ) \tilde{p}_\alpha \tilde{p}_\alpha + (2\delta_{\nu,1} -1 ) p_1 q - \minHs^2}{(\nu_n - \ii \muSh)^2 + (\mathbf{p} + \mathbf{q})^2 + \minHs^2} \frac{1}{(\nu_n - \ii \muSh)^2 + \mathbf{p}^2 + \minHs^2}, \label{eq:H_wn_wn}
\end{align}
\end{widetext}
where we choose the angle integration such that $q$ lies on the $x_1$ axis.
The off-diagonal elements are given by 
\begin{widetext}
\begin{align}
	H_{\sigma \omega_\nu} &= \left(\delta_{\nu,3} \left(-\ii -1 \right) + 1 \right) \minHs \, \frac{N_\gamma}{\beta} \int \frac{\dr^2 p}{(2\pi)^2} \sum_{n=-\infty}^{\infty} \frac{2 \tilde{p}_\nu + \delta_{\nu,1} p_1 q}{(\nu_n - \ii \muSh)^2 + (\mathbf{p} + \mathbf{q})^2 + \minHs^2} \frac{1}{(\nu_n - \ii \muSh)^2 + \mathbf{p}^2 + \minHs^2}, \label{eq:H_s_wn} \\
	H_{\omega_1 \omega_2} &= \frac{N_\gamma}{\beta} \int \frac{\dr^2 p}{(2\pi)^2} \sum_{n=-\infty}^{\infty} \frac{2 p_1 p_2 + q p_2}{(\nu_n - \ii \muSh)^2 + (\mathbf{p} + \mathbf{q})^2 + \minHs^2} \frac{1}{(\nu_n - \ii \muSh)^2 + \mathbf{p}^2 + \minHs^2}, \label{eq:H_w1_w2}  \\
	H_{\omega_3 \omega_j} &= -\frac{N_\gamma}{\beta} \int \frac{\dr^2 p}{(2\pi)^2} \sum_{n=-\infty}^{\infty} \frac{2 \tilde{p}_3 p_j + \delta_{j,1} \tilde{p}_3 q}{(\nu_n - \ii \muSh)^2 + (\mathbf{p} + \mathbf{q})^2 + \minHs^2} \frac{1}{(\nu_n - \ii \muSh)^2 + \mathbf{p}^2 + \minHs^2}, \quad j \in \{1,2\}. \label{eq:H_w3_wj}\\
\end{align}
\end{widetext}
Note that in consistency with the repulsive nature of the Yukawa interaction $\bar \psi \omega_3 \psi$, the Hessian matrix elements with only one $\omega_3$ index are purely imaginary.  

\subsection{Finite temperature expressions for non-vanishing q}
In order to evaluate the Hessian at finite temperature and chemical potential, we perform the Matsubara summation. 
For some of the above expression, one needs to do some additional manipulations in order to perform the summation in an easier way or to circumvent problems in the contour integration stemming from $\tilde{p}_3^2$ terms in the numerator.
Equation \eqref{eq:H_ss} is exactly the two-point vertex function of the \gls{gn} model, and was calculated multiple times, e.g., in \Rcite{Buballa:2020nsi, Pannullo:2023cat, Koenigstein:2023yzv}, see Appendix B of \Rcite{Pannullo:2023one} for the evaluation of the two-point vertex function in various limits of $\minHs$, $T$ and $q$.
The other entries, however, have to our knowledge not be computed. 
Using some manipulation of the numerator of \cref{eq:H_wn_wn} for $\nu = 3$ as well as rewriting the denominator in a partial fraction (see Eqs.~$(4.14-16)$ in \Rcite{Klevansky:1992qe} for the general idea) one finds that
\begin{align}
H_{\omega_3 \omega_3} = \frac{1}{\lambda_V} &- \ell_1(\mu, T, \minHs, \minHw) + L_{2, +}(\muSh, T, \minHs, q) + \nonumber \\ &+ \ell_3(\mu,T,\minHs, q), \label{eq:Hw3w3eval}
\end{align}
where $\ell_1$ is known from the \gls{gn} model and appears in the gap equation \eqref{eq:GapSigma}, 
\begin{widetext}
\begin{equation}
	L_{2, +}(\muSh, T, \minHs, q) = \frac{1}{2} \int \frac{\dr^2 p}{(2\pi)^2} \sum_{n=-\infty}^{\infty} \frac{N_\gamma}{\beta} \frac{\left(q^2 + 4 \minHs^2\right)}{(\nu_n - \ii \muSh)^2 + \mathbf{p}^2 + \minHs^2} \frac{1}{(\nu_n - \ii \muSh)^2 + \left(\mathbf{p} + \mathbf{q}\right)^2 + \minHs^2} \equiv \frac{1}{2}\left(q^2 + 4 \minHs^2\right)  \ell_2(\mu, T, \minHs, q)
\end{equation}
\end{widetext}
is (up to a factor of $2$) the momentum dependence of the two-point vertex function in the \gls{gn} model (compare Eq.~(13) of \Rcite{Pannullo:2023one}) and the new contribution 
\begin{align}
\ell_3(\muSh,T,\minHs, q) = \tfrac{2 N_\gamma}{\beta} &\int \frac{\dr^2 p}{(2\pi)^2} \nonumber \times \\ \sum_{n=-\infty}^{\infty}  \frac{\mathbf{p}^2 + \mathbf{p} \mathbf{q}}{(\nu_n - \ii \muSh)^2 + \mathbf{p}^2 + \minHs^2}	&\frac{1}{(\nu_n - \ii \muSh)^2 + \left(\mathbf{p} + \mathbf{q}\right)^2 + \minHs^2}.
\end{align}
The Matsubara summation can be performed as usual by analytic continuation and a contour integral. 
Then, one obtains 
\begin{align}
\ell_3(\muSh,T,\minHs, q) = 2 N_\gamma  \int \frac{\dr^2 p}{(2\pi)^2} \frac{\left(p^2 + \mathbf{p} \mathbf{q}\right)}{q^2 + 2\mathbf{p} \mathbf{q}} \left[\frac{1}{2 E}\left(1 - \FD(E) - \AFD(E)\right) - \left(E \rightarrow E_q\right) \right]&
\end{align}
with $E$, $\FD(x)$, $\AFD(x)$ as defined in \cref{sec:methomcond} and $E_q = \sqrt{(p + q)^2 + \minHs^2}$.
This expression can be split up between $p_1 < q /2$ and $p_1 > q/2$, manipulated and evaluated using a Cauchy-Principal value (similar to how $L_2$ can be evaluated) and amounts to
\begin{widetext}
\begin{equation}
	\ell_3(\muSh,T,\minHs, q) = \frac{N_\gamma}{(4\pi)} \left\{\int_0^{q/2} \dr p\,\frac{p}{E} \frac{p^2 - q^2 /2 }{q \sqrt{q^2 - 4p}} \left[1 - \FD(E) - \AFD(E)\right] + \int_0^\infty \frac{p}{E} \left[1 - \FD(E) - \AFD(E)\right] \right\},
\end{equation}
\end{widetext}
where the latter term is linearily divergent and identical to $\ell_1$. 
Thus, the divergences of $\ell_1$ and $\ell_3$ exactly cancel each other out such that \cref{eq:Hw3w3eval} is finite. 
We are aware that these divergences occur because of splitting up the numerator of \cref{eq:H_wn_wn} for $\nu = 3$, as described above.
However, this procedure is easier than dealing with the $\left(\nu_n - \ii \mu\right)^2$ term in the numerator, since the standard method of analytic continuation and contour integration cannot be done as usual, since the integrand does not fall off fast enough.

In turn, the other diagonal elements are rather straight forward
\begin{widetext}
\begin{equation}
	H_{\omega_j \omega_j} = \frac{1}{\lambda_V} - \ell_1 + \frac{1}{2} q^2 \ell_2 + \frac{2 N_\gamma}{\beta} \int \frac{\dr^2 p}{(2\pi)^2} \sum_{n=-\infty}^{\infty} \frac{p_j^2 + \delta_{j,1} p_1 q}{(\nu_n - \ii \muSh)^2 + \mathbf{p}^2 + \minHs^2} \frac{1}{(\nu_n - \ii \muSh)^2 + \left(\mathbf{p} + \mathbf{q}\right)^2 + \minHs^2}.
\end{equation}
\end{widetext}
Splitting up the numerator and treating the remaining integrals with standard techniques (shifts, inversions of the integration variable) allow to identify already known integral structures for $j = 1$ and one obtains
\begin{align}
H_{\omega_1 \omega_1} &= \frac{1}{\lambda_V} - \ell_1 + \frac{1}{2} q^2 \ell_2 + \ell_1 + q^2 \ell_2 \nonumber \\  &=  \frac{1}{\lambda_V} + \frac{3}{2} q^2 \ell_2(\mu, T, \minHs, q).
\end{align}
Also, one finds for $j=2$
\begin{align}
H_{\omega_2 \omega_2} &= \frac{1}{\lambda_V} +  \nonumber \\ + \frac{N_\gamma}{2\pi q} \int_0^{q/2} & \frac{p^3}{E \sqrt{q^2 - 4 p^2}} \left[1 - \FD(E) - \AFD(E)\right],
\end{align}
which is again finite. 
Note that the differences in the expression for $H_{\omega_1 \omega_1}$ and $H_{\omega_1 \omega_1}$ come from the choice on $\mathbf{q} = \left(q, 0\right)$.
The matrix elements $H_{\omega_2 \omega_2}$ and $H_{\omega_1 \omega_1}$ would be exchanged if we would have chosen $\mathbf{q} = \left(0, q\right)$, as expected since the analysis is invariant under spatial rotations.
The limits of zero temperature in the above expressions is rather straightforward. 

The mixing between $\sigma$ and $\omega_3$ is given by 
\begin{align}
H_{\sigma \omega_3} = -\ii \minHs\frac{2N_\gamma}{\beta} &\int \frac{\dr^2 p}{(2\pi)^2} \sum_{n=-\infty}^{\infty} \frac{\tilde{p}_3}{(\nu_n - \ii \muSh)^2 + \mathbf{p}^2 + \minHs^2} \times \nonumber \\ &\times  \frac{1}{(\nu_n - \ii \muSh)^2 + \left(\mathbf{p} + \mathbf{q}\right)^2 + \minHs^2}.
\end{align}
This expression can be evaluated using a contour integral, as the analytic continuation of the integrand is well-behaved when closing the contour at infinity.
The result yields
\begin{widetext}
\begin{align}
	H_{\sigma \omega_3}(q) &= \ii \minHs \N_\gamma \int \frac{\dr^2 p}{(2\pi)^2} \frac{1}{\mathbf{q}^2 + 2 \mathbf{p} \mathbf{q}} \left[\left(\FD(E) - \AFD(E)\right) - \left(\FD(E_q) - \AFD(E_q)\right)\right] \\  &= -\ii \minHs \N_\gamma \int_0^{q/2} \frac{\dr p}{2\pi} \frac{p}{q \sqrt{q^2-4p^2}} \left[\FD(E) - \AFD(E)\right] \equiv -\ii \minHs \ell_4(\mu,T, \minHs, q) \nonumber,
\end{align}
\end{widetext}

where one can directly see that this expression vanishes in the \gls{sp} and for $\mu = 0$.

Also, one obtains 
\begin{equation}
H_{\sigma \omega_1} = 2 \minHs |q| \ell_2(\mu, T, \minHs, q) \label{eq:H_s_w1eval}
\end{equation}
and $H_{\sigma \omega_2}=0$, where the term with $j=2$ vanishes using symmetry arguments in the angle integration\footnote{At first glance, this seems to be a trivial integration, but one still needs to perform a Cauchy principal value before using symmetry arguments.} (compare \cref{eq:H_s_wn} for $\nu=2$). 
Again, the two matrix elements $H_{\sigma \omega_1}$ and $H_{\sigma \omega_1}$ would be exchanged, if we would setup the momentum integration such that $\mathbf{q}$ would be aligned with the $p_2$ axis.
In a similar way as for $H_{\sigma \omega_2}$, we obtain that $H_{\omega_1 \omega_2} = H_{\omega_3 \omega_2} = 0$.
Lastly, one obtains
\begin{equation}
H_{\omega_3 \omega_1} = 0 
\end{equation}
through splitting up the integral in \cref{eq:H_w3_wj} into two integrals, where each integral contains one of the summands in the numerator.
Both integrals turn to be proportional to $\ell_4$, but with different signs and prefactors such that both contributions cancel each other.
\subsection{Finite temperature expressions for $q=0$\label{app:zeroqStab}}
For the $q=0$ expressions (and even more limits) of $H_{\sigma \sigma}$ and all integrals related to it, we refer again to \Rcite{Pannullo:2022eqh}.
Note that several of the integrals in the beginning of \cref{app:stab_analysis} vanish, when the limit of $q$ going to zero is taken.
More precisely, we find
\begin{align}
H_{\sigma \omega_j}(q=0) &= 0, \quad H_{\omega_3 \omega_j}(q=0)= 0 , \quad j \in \{1,2\}, \\
H_{\omega_1 \omega_2}(q=0) &= 0,
\end{align}
as can be seen from going into polar coordinates after Matsubara summation and performing the angle integration, respectively.
The Hessian \eqref{eq:Hessian} becomes diagonal w.r.t.~the $2\times2$ block with $\phi_j, \phi_k \in \{\omega_1, \omega_2\}$.
Thus, complex-conjugate eigenvalue pairs can only be generated by mixing of $\sigma$ and $\omega_3$ at $q=0$, allowing to study this phenomenon by only taking the respective $2\times 2$ block in \cref{eq:Hessian}. 
For the rest of the diagonal elements, we find 
\begin{widetext}
\begin{align}
	H_{\omega_j \omega_j}(q=0) &= \frac{1}{\lambda_V} - \ell_1(\mu, T, \minHs) + \ell_3(\mu, T, \minHs, q = 0)  \\ &= \frac{1}{\lambda_V} - \ell_1 + \frac{N_\gamma}{8\pi} \int_0^\infty \dr p \frac{p^3}{E^3} \left[1 - \left(1 + \beta E\right) \left(\FD(E) + \AFD(E)\right) + \beta E \left(\FD^2(E) + \AFD^2(E)\right)\right] \\ &= \nonumber \frac{1}{\lambda_V} + \frac{N_\gamma}{8\pi} \int_{|\minHs|}^\infty \dr E\, \Big\{\beta E \left[ \FD^2(E) + \AFD^2(E) - \FD(E) - \AFD(E)\right] \Big. + \\ &\hspace{1cm}- \Big. \frac{\minHs^2}{E^2} \left[ 1 - \left(1 + \beta E\right) \left(\FD(E) + \AFD(E)\right) + \beta E \left(\FD^2(E) + \AFD^2(E)\right)\right]\Big\}   \nonumber,
\end{align}
\end{widetext}

which is finite due to the suppression of high momenta in the first summand of the integrand, and 
\begin{align}
H_{\omega_3 \omega_3}(q=0) = \frac{1}{\lambda_V} - \ell_1 + 2 \minHs^2 \ell_2 + \ell_3,
\end{align}
which is identical to $H_{\omega_j \omega_j}$ up to an additional contribution of $2\minHs^2 \ell_2$. 
The diagonal element $H_{\sigma \sigma}$ is given by the bosonic two-point vertex function $\Gamma^{(2)}_\sigma(q)$ in the $\left(2+1\right)$-dimensional \gls{gn} model and studied in this limit in \Rcite{Buballa:2020nsi,Pannullo:2023one}.
For the remaining off-diagonal element we find 
\begin{align}
H_{\sigma \omega_3} &= -\ii \minHs \frac{N_\gamma}{\pi} \times \nonumber \\ \times \int_0^\infty \dr E &\left\{ \beta E \left[ \FD^2(E) - \AFD^2(E)\right] - \left[ \FD(E) - \AFD(E)\right]\right\}.
\end{align}

\bibliography{literature}
\end{document}